\documentclass{SciPost}

\hypersetup{
    colorlinks,
    linkcolor={red!50!black},
    citecolor={blue!50!black},
    urlcolor={blue!80!black}
}

\usepackage[bitstream-charter]{mathdesign}
\renewcommand{\pi}{\piup}
\DeclareSymbolFont{usualmathcal}{OMS}{cmsy}{m}{n}
\DeclareSymbolFontAlphabet{\mathcal}{usualmathcal}

\fancypagestyle{SPstyle}{
\fancyhf{}
\lhead{\colorbox{scipostblue}{\bf \color{white} ~SciPost Physics}}
\rhead{{\bf \color{scipostdeepblue} ~Submission }}

\fancyfoot[C]{\textbf{\thepage}}
}

\usepackage[T1]{fontenc} 
\usepackage{mathtools,amsthm,physics-patch,dsfont}
\usepackage{url}
\usepackage[shortlabels]{enumitem}
\setlist[enumerate,1]{1),itemsep=2pt,parsep=0pt,topsep=3pt,left=0pt}
\setlist[itemize,1]{left=0pt}
\usepackage{float}
\usepackage{tikz}
\usetikzlibrary{decorations.markings}
\newcommand{\Arccos}{\operatorname{Arccos}}
\newcommand{\De}{\Delta}
\renewcommand{\Im}{\operatorname{Im}}
\newcommand{\Li}{\operatorname{Li}}
\newcommand{\T}{\mathsf T}
\newcommand{\Th}{\Theta}
\newcommand{\ZZ}{{\mathbb Z}}
\newcommand{\all}{\forall}
\newcommand{\bH}{\mathbf H}
\newcommand{\al}{\alpha}
\newcommand{\de}{\delta}

\newcommand{\diag}{\operatorname{diag}}
\newcommand{\e}{\mathrm e}

\renewcommand{\ge}{\geqslant}
\renewcommand{\geq}{\geqslant}
\newcommand{\iu}{\mathrm i}
\renewcommand{\le}{\leqslant}
\renewcommand{\leq}{\leqslant}
\newcommand{\pd}{\partial}
\newcommand{\sgn}{\operatorname{sgn}}
\newcommand{\si}{\sigma}

\renewcommand{\th}{\theta}
\newcommand{\vep}{\varepsilon}

\newcommand{\vp}{\varphi}

\newcommand{\hc}{\mathrm{H.c.}}
\newcommand{\vac}{\text{vac}}
\let\br\PTqty
\newcommand{\Cl}{\operatorname{Cl}}
\theoremstyle{remark}
\newtheorem{remark}{Remark}[section]

\begin{document}

\pagestyle{SPstyle}

\begin{center}{\Large \textbf{\color{scipostdeepblue}{
Local fermion density in inhomogeneous free-fermion chains:\\
a discrete WKB approach\\
}}}\end{center}

\begin{center}\textbf{
Martín Zapata\textsuperscript{1$\star$},
Federico Finkel\textsuperscript{2$\dagger$} and
Artemio González-López\textsuperscript{2$\ddagger$}
}\end{center}

\begin{center}
  {\bf 1} Theory Division, Max Planck Institute of Quantum Optics, D-85748 Garching,  Germany\\
  {\bf 2} Departamento de Física Teórica, Facultad de Ciencias Físicas, Universidad Complutense
  de Madrid, Plaza de las Ciencias 1, 28040 Madrid, Spain
  \\[\baselineskip]
  $\star$ \href{mailto:martin.zapata@mpq.mpg.de}{\small martin.zapata@mpq.mpg.de}\,,\quad
  $\dagger$ \href{mailto:ffinkel@ucm.es}{\small ffinkel@ucm.es}\,,\quad $\ddagger$
  \href{mailto:artemio@ucm.es}{\small artemio@ucm.es}
\end{center}

\section*{\color{scipostdeepblue}{Abstract}}
\textbf{\boldmath{%
    We introduce a novel analytical approach for studying free-fermion (XX) chains with smoothly
    varying, site-dependent hoppings and magnetic fields. Building on a discrete WKB-like
    approximation applied directly to the recurrence relation for the single-particle
    eigenfunctions, we derive a closed-form expression for the local fermion density profile as a
    function of the Fermi energy, which is valid for arbitrary fillings, hopping amplitudes and
    magnetic fields. This formula reproduces the depletion and saturation effects observed in
    previous studies of inhomogeneous free-fermion chains, and provides a theoretical framework to
    understand entanglement entropy suppression in these models. We demonstrate the accuracy of
    our asymptotic formula in several chains with different hopping and magnetic field profiles.
    Our findings are thus the first step towards an analytical treatment of entanglement in
    free-fermion chains beyond the reach of conventional field-theoretic techniques.
}}

\vspace{\baselineskip}

\noindent\textcolor{white!90!black}{%
\fbox{\parbox{0.975\linewidth}{%
\textcolor{white!40!black}{\begin{tabular}{lr}%
  \begin{minipage}{0.6\textwidth}%
    {\small Copyright attribution to authors. \newline
      This work is a submission to SciPost Physics. \newline
      License information to appear upon publication. \newline
      Publication information to appear upon publication.}
  \end{minipage} & \begin{minipage}{0.4\textwidth}
    {\small Received Date \newline Accepted Date \newline Published Date}%
  \end{minipage}
\end{tabular}}
}}
}


\vspace{10pt}
\noindent\rule{\textwidth}{1pt}
\tableofcontents
\noindent\rule{\textwidth}{1pt}
\vspace{10pt}



\section{Introduction}\label{sec:intro}

The study of one-dimensional quantum systems is central to our understanding of many-body physics,
particularly in the context of condensed matter physics and quantum entanglement. In recent years,
attention has increasingly focused on inhomogeneous spin chains, in which hopping amplitudes and
external magnetic fields vary across the lattice, due to their relevance in cold atomic systems on
optical lattices~\cite{GL14,BMF21}, engineered quantum simulators~\cite{RTLC17,SFSTT20,MCDetal21},
and the study of entanglement in non-uniform environments~\cite{VRL10,RDRCS17,TRS18}. Among these
chains, the inhomogeneous version of the spin-$\frac{1}{2}$ Heisenberg XX chain~\cite{LSM61}
occupies a prominent place, both for its conceptual simplicity and its connections to free fermion
systems. With open boundary conditions, and in the presence of an external magnetic field, the
Hamiltonian of this model can be taken as\footnote{As shown in Ref.~\cite{FG20}, all the models
  obtained replacing $J_n$ by $\vep_n J_n$, where $\vep_n\in\{\pm1\}$ is a site dependent sign,
  are actually unitarily equivalent.}
\begin{equation}\label{eq:HXX}
  H =
  \frac{1}{2}\sum_{n=0}^{N-2}J_n(\sigma^x_n\sigma^x_{n+1}+\sigma^y_n\sigma^y_{n+1})
  +\frac{1}{2}\sum_{n=0}^{N-1}B_n(1-\sigma^z_n)\,,
\end{equation}
where $\si_n^\al$ ($\al=x,y,z$) denotes the Pauli matrix $\si^\al$ acting on the $n$-th site. As
is well known, this Hamiltonian can be mapped to the fermionic Hamiltonian
\begin{equation}
  H = \sum_{n=0}^{N-2}J_n(c^\dagger_nc^{}_{n+1}+\hc)+\sum_{n=0}^{N-1}B_nc^\dagger_n c^{}_n
  \label{eq:fermion_Hamiltonian}
\end{equation}
through the Jordan--Wigner transformation~\cite{JW28}
\begin{equation}\label{eq:JW}
  c_n\mapsto\prod_{j=0}^{n-1}\si_j^z\cdot \si_n^{+},
  \qquad\si_n^\pm\coloneqq\frac12(\si_n^x\pm\iu\si_n^y).
\end{equation}
Since $c_n^\dagger$ creates a (spinless) fermion at site $n$, the
Hamiltonian~\eqref{eq:fermion_Hamiltonian} describes a system of hopping fermions with hopping
amplitude $J_n$ and chemical potential $B_n$. As explained in the next section, the full spectrum
of the Hamiltonian~\eqref{eq:fermion_Hamiltonian} consists of excitations of single-particle
energy modes. Moreover, since the Hamiltonian~\eqref{eq:fermion_Hamiltonian} conserves particle
number, it is possible to construct a basis of energy eigenstates with well-defined total fermion
number $M$ and filling fraction $\nu=M/N\in[0, 1]$.

The XX model is one of the few systems for which the (bipartite) entanglement entropy of energy
eigenstates can be numerically evaluated for a large number of spins~\cite{VLRK03,Pe03,LR09}. In
general, given a bipartition of a quantum system into two disjoint blocks $A$ and $B$, the
bipartite entanglement entropy $S$ of pure state $\ket\psi$ of the whole system is defined by
\begin{equation}
  S=s[\rho_A]\,,\qquad
  \rho_A=\tr_{B}(\rho)\,,
\end{equation}
where $\rho=\dyad\psi$ is the density matrix of the state~$\ket\psi$, $\rho_A$ is the (in general,
mixed) reduced density matrix of subsystem $A$, and $s$ is an entropy functional (usually the
Rényi entropy~\cite{Re70}). A particularly intriguing phenomenon observed in inhomogeneous XX
chains is the suppression of bipartite entanglement entropy at certain block sizes for fillings
$\nu\neq1/2$ and/or inhomogeneous magnetic fields~\cite{FG21}, in contrast to the behavior
observed in the homogeneous case ($J_n=J,B_n=B$). Mula et al.~\cite{MSSSR22} provided a natural
explanation of this phenomenon in terms of an exponential reduction, termed `depletion', of the
local fermionic density. Using field-theoretic techniques, these authors derived an approximate
heuristic formula for the occupation profile, valid only in the low-filling limit \emph{and} for
zero magnetic field. This formula was able to account for depletion in this regime, and in
particular to predict the location of the depletion intervals in several simple examples.
Nevertheless, a general, analytically tractable approach valid for arbitrary fillings/magnetic
fields has so far remained elusive.

In this paper we introduce a novel method for studying the local density of inhomogeneous
free-fermion chains in the thermodynamic limit. Our main idea is to directly analyze the behavior
of the model's single-particle eigenfunctions in the large $N$ limit, avoiding intermediate
approximations. In particular, we do not base our analysis (as is usually done in the literature;
see, e.g., refs.~\cite{RRS15,RDRCS17,TRS18,MSSSR22}) on field-theoretic techniques, which often
rely on approximations valid only in the absence of a magnetic field, and either near the critical
regime (half filling) or at low filling. Our approach is rather based on applying a WKB-like
approximation to the three-term recurrence relation satisfied by the single-particle
eigenfunctions of the discrete model in the large $N$ limit. In this way we obtain continuous
(`WKB') approximations to these eigenfunctions, which we then use to derive an asymptotic
expression for the fermionic density as a function of the Fermi energy $\vep_F$ valid for
arbitrary filling fractions and/or external magnetic fields. More precisely, we show that the
average occupation number at an arbitrary Fermi energy $\vep_F$, hopping $J_n$ and magnetic field
$B_n$ is approximately given by
\begin{equation}
  \label{eq:occ_dens}
  \expval*{c^\dagger_nc_n}\simeq
  \begin{cases}
      0,&\vep_F\leq B_n-2J_n\,,\\[1mm]
      \dfrac1\pi\arccos(\dfrac{B_n-\vep_F}{2J_n}),\qquad & B_n-2J_n\le\vep_F\le B_n+2J_n\,,\\[1mm]
      1,& \vep_F\geq B_n+2J_n\,.
    \end{cases}
  \end{equation}
In the absence of an external magnetic field our result agrees with ref.~\cite{MSSSR22} in the low
filling limit, while at high filling we find a mirror image of depletion, which we may call
`saturation', that the heuristic formula in the last reference cannot reproduce. In particular,
eq.~\eqref{eq:occ_dens} is able to accurately determine the saturation regions, which can appear
even for relatively low values of the filling fraction in sufficiently strong magnetic fields. We
validate our analytical predictions through extensive numerical simulations across several classes
of inhomogeneous chains, including the rainbow, Krawtchouk, cosine, and Rindler\footnote{For the
  Rindler chain, see the Supplementary Material.} chains. In all cases, eq.~\eqref{eq:occ_dens} is
in excellent agreement with the numerical results over the whole range of energies/filling
fractions, hoppings and magnetic fields involved. Beyond its immediate applicability to fermion
density profiles, our method is also a promising starting point for deriving asymptotic formulas
for correlation matrices and entanglement entropies in inhomogeneous fermionic chains,
particularly where conformal field theory methods are not directly applicable due to the lack of
scale invariance.

Our paper is structured as follows. In section~\ref{sec:prelim} we collect several well-known
preliminary results used throughout the paper. In particular, we recall the role of
single-particle states in the exact diagonalization of the model, and its connections with the
theory of orthogonal polynomials. In section~\ref{sec:WKB} we apply the discrete WKB approximation
to the recurrence relation satisfied by the single-particle eigenfunctions. We derive approximate
WKB wave functions, which are then used in section~\ref{sec:spectrum} to obtain accurate
asymptotic formulas for the fermionic density profile and the filling fraction in the
thermodynamic limit. In section~\ref{sec:examples} we apply our method to several well-known types
of inhomogeneous XX chains like the Krawtchouk, rainbow, and cosine chains, including an
asymmetric generalization of the latter. Finally, in section~\ref{sec:conc} we summarize the
paper's main results and discuss natural directions for future work suggested by our findings.

\section{Preliminaries}\label{sec:prelim}

Let us first briefly recall how to diagonalize the fermionic
Hamiltonian~\eqref{eq:fermion_Hamiltonian}.
To this end, note that this Hamiltonian is a quadratic function of fermionic operators:
\begin{equation}
  H=\sum\limits_{n,m=0}^{N-1}H_{nm}c^\dagger_nc_m,\qquad
  H_{nm}\coloneqq J_n\delta_{m,n+1}+J_{n-1}\delta_{m,n-1}+B_n\delta_{nm}\,,
  \label{eq:tridiag}
\end{equation}
where $J_{-1}=J_{N-1}=0$. Using vector notation
\[
  \vb{c}=(c_1,\dots,c_N)^\T\,,\quad \vb{c^\dagger}=(c_1^\dagger,\dots,c_N^\dagger)\,, \quad
  \vb{H}=(H_{nm}),
\]
we can more concisely write
\[
  H=\vb{c}^\dagger\vb{H}\,\vb{c}.
\]
Let $\ket{n}=c_n^\dagger\ket{\vac}$ denote the state with exactly one fermion at site $n$,
so that
\[
  \{\ket n:n=0,\dots,N-1\}
\]
is an orthonormal basis of the single-particle subspace. We then have
\[
  H_{nm}=\mel{n}Hm,
\]
and thus $\vb{H}$ is the matrix of the restriction of $H$ to the single-particle sector.

Since the matrix $\vb{H}$ is real and symmetric, it can be diagonalized by means of a suitable
real orthogonal matrix $\vb{\Phi}=(\Phi_{nk})$. More precisely, setting
\[
  \tilde{\vb{c}}=\vb{\Phi}^\T\vb{c}
\]
we have
\begin{equation}
  H = \tilde{\vb{c}}^\dagger\tilde{\vb{H}}\,\tilde{\vb{c}}\,,\qquad\text{with}\quad
  \tilde{\vb{H}}=\vb{\Phi}^\T\vb{H}\,\vb{\Phi}=\diag(\varepsilon_0,\dots,\varepsilon_{N-1})\,,
  \label{eq:vector}
\end{equation}
where $\vep_0<\cdots<\vep_{N-1}$ are the eigenvalues of $\bH$---i.e., the single-particle
energies. The new operators are still fermionic by virtue of the orthogonality of $\vb{\Phi}$. The
operator $\tilde c_k^\dagger$ creates a single fermion with well-defined energy $\vep_k$, i.e., a
single-particle energy mode. Since, by eq.~\eqref{eq:vector},
\begin{equation}\label{eq:Hdiag}
  H=\sum_{k=0}^{N-1}\vep_k\tilde c^\dagger_k\tilde c_k,
\end{equation}
the eigenstates of
$H$ can therefore be constructed by applying the operators
$\tilde{c}^\dagger_k$ to the vacuum state $\ket{\vac}$, annihilated by all
$\tilde{c}_k$. In other words, the states
\begin{equation}
  \ket{n_0,n_1,\dots,n_{k}}\coloneqq
  \tilde{c}^\dagger_{n_0}\tilde{c}^\dagger_{n_1}\cdots\tilde{c}^\dagger_{n_k}
  \ket{\vac}\,,\qquad 0\le n_0<\cdots<n_k\le N-1,
\end{equation}
are eigenstates of $H$ with energies
\begin{equation}
  E(n_0,n_1,\dots,n_k) = \sum_{j=0}^{k}\varepsilon_{n_j},
\end{equation}
and they form an orthonormal basis of the Hilbert space of $H$. Thus the full spectrum of $H$ is
obtained by exciting a number of single-particle energy modes.

It is well known that the eigenvectors
$\phi_n(\varepsilon_k)\coloneqq\Phi_{nk}$ are related to the family of monic orthogonal polynomials
$\{P_n(\varepsilon)\}_{n=0}^{N}$ defined recursively via
\begin{equation*}
  P_{n+1}(\varepsilon)=(\varepsilon-b_n)P_n(\varepsilon)-a_{n-1}P_{n-1}(\varepsilon)\,,
  \qquad\text{with}\quad
  P_{-1}(\vep)=0,\quad P_0(\varepsilon)=1\,,
\end{equation*}
where
\[
  a_n=J_n^2,\qquad b_n=B_n
\]
(see, e.g.~\cite{CNV19, FG20, FG21}). Specifically,
$\phi_n(\varepsilon_k)\propto P_n(\varepsilon_k)$, and the spectrum of the single-particle
Hamiltonian coincides with the set of roots of the \emph{critical polynomial} $P_N$. This
equivalence has, for instance, aided in the classification of inhomogeneous XX chains of different
types \cite{FG20} and provided some exact spectra \cite{CNV19,FG21,BPV24,BCLMV25}. It also
guarantees that the single-particle eigenstates are non-degenerate.

If $M=0,\dots,N$, we define the $M$-filled state $\ket M$ as the state obtained by exciting the
lowest $M$ single-particle energy modes:
\begin{equation}
  \ket{M}=\prod_{k=0}^{M-1}\tilde{c}^\dagger_k\ket{\vac}\,.
  \label{eq:ground_state}
\end{equation}
In particular, the system's ground state is the state $\ket M$ for which $\vep_i<0$ for
$i=0,\dots,M-1$ and $\vep_M>0$ (of course, the ground state is twice degenerate if there happens
to be a zero mode). The correlation matrix $\vb{C}=(C_{nm})$ for this state $\ket M$ is defined by
\begin{equation}
  C_{nm}\coloneqq\mel{M}{c^\dagger_nc^{}_m}{M}=\sum_{k=0}^{M-1}\Phi_{nk}\Phi_{mk}\,,
  \label{eq:correlation_matrix}
\end{equation}
where the last equality follows from the identity
\begin{equation*}
  \mel{M}{\tilde{c}^\dagger_k\tilde{c}_q}{M}=\delta_{kq}\chi_M(k),
\end{equation*}
with
\begin{equation*}
  \chi_M(k)=
  \begin{cases}
    1,\qquad k\in\{0,\dots,M-1\}\,,\\
    0,\qquad k\not\in\{0,\dots,M-1\}\,.
  \end{cases}
\end{equation*}
As is well known~\cite{VLRK03,Pe03,LR09}, when the chain is in one of the states $\ket M$ defined
above the entanglement entropy $S_A$ of a block $A$ of $L$ spins can be expressed in terms of the
eigenvalues $\lambda_l$ (with $l=0,\dots,L-1$) of the $L\times L$ truncated correlation matrix
$\vb C_A=(C_{nm})_{n,m\in A}$ through the formula\footnote{More precisely, the following formula
  is valid if the entropy functional $S$ is additive, as is the case for the Rényi and von Neumann
  entropies.}
\begin{equation}
  S_A=\sum_{l=0}^{L-1}s^{(2)}(\lambda_l)\,,
  \label{eq:entropy}
\end{equation}
where $s^{(2)}$ is the binary entropy associated to the entropy functional $S$.

As mentioned in the introduction, in this work we shall be mainly concerned with the diagonal
elements of the (full) correlation matrix
\begin{equation}
  \label{eq:Cnn}
  C_{nn}=\mel{M}{c^\dagger_nc^{}_n}{M}
  =\sum_{k=0}^{M-1}\Phi_{nk}^2\eqqcolon\expval*{c_n^\dagger c_n},\qquad n=0,\dots,N-1,
\end{equation}
which provide the expectation value of the $n$-th site occupation number when the system is in the
$M$-filled state $\ket M$. Although in this paper we shall not deal with the entanglement entropy,
it is perhaps of interest to briefly explain how the entanglement entropy depletion arises from
the depletion/saturation of occupation. To this end, note that since
\[
  0=\expval*{c^\dagger_nc_n}=\norm\big{c_n\ket{M}}^2\implies
  c_n\ket{M}=0,
\]
if (for instance) $\expval*{c^\dagger_nc^{}_n}$ vanishes for $n=0,\dots,L-1$ the state $\ket M$ is
the tensor product of the vacuum $\otimes_{n=0}^{L-1}\ket{\text{vac}}_n$ at sites $0,\dots,L-1$
with a pure state in the Hilbert space of the remaining sites. As a consequence, the entanglement
entropy of a block $\{0,\dots,r-1\}$ with $r\le L$ spins vanishes. The same conclusion holds
if~$\expval*{c^\dagger_nc_n}=1$ for $n=0,\dots,L-1$, since
\[
  1=\expval*{c^\dagger_nc^{}_n}=1-\expval*{c_nc^\dagger_n}\implies
  0=\expval*{c_nc^\dagger_n}=\norm\big{c_n^\dagger\ket{M}}^2\implies c_n^\dagger\ket{M}=0,
\]
and hence in this case $\ket M$ is the tensor product of the fully occupied state
$\otimes_{n=0}^{L-1}\ket{f}_n$ with a pure state in the Hilbert space of the sites $L,\dots,N-1$.

\section{Single-particle WKB wave functions}\label{sec:WKB}

While the properties discussed in the previous section are valid in general, when trying to derive
analytical results we will concern ourselves with a limited class of chains, with slowly varying
hoppings and external magnetic field strength. More concretely, we introduce an arbitrarily small
lattice spacing $a$ and define the variable $x=na\in[0,\ell]$, such that the chain's length
$\ell=Na$ remains finite as $a\rightarrow0$ and $N\to\infty$. In this way, in the thermodynamic
limit we can regard $x$ as a continuous coordinate along the chain. We will assume that $J_n$ and
$B_n$ converge pointwise to sufficiently smooth functions
\begin{equation}
  J_n\underset{a\to0}\longrightarrow J(x)\,,\quad
  B_n\underset{a\to0}\longrightarrow  B(x)\,,
\end{equation}
where our notion of pointwise convergence for lattice functions $f_{n;N}$ will be
\begin{equation}
  \lim_{\substack{N\rightarrow\infty\\na\rightarrow x}}
  f_{n;N}=f(x)\,.
\end{equation}

The eigenvalue problem for the $N\times N$ real symmetric matrix $\bH$,
\begin{equation*}
  \sum_{m=0}^{N-1}H_{nm}\phi_m(\vep_k)=\vep_k\phi_n(\vep_k),\qquad n=0,\dots,N-1,
\end{equation*}
can be written as
\begin{equation}
  \phi_{n+1}(\vep_k)+\frac{J_{n-1}}{J_n}\phi_{n-1}(\vep_k)=2\xi_n(\vep_k)\phi_n(\vep_k)\,,\qquad
  \xi_n(\vep_k)\coloneqq\frac{\vep_k-B_n}{2J_n}\,,
  \label{eq:recurrence:phi}
\end{equation}
subject to Dirichlet boundary conditions $\phi_{-1}(\vep_k)=\phi_{N}(\vep_k)=0$. To recast these
equations into a more symmetric form, we begin by defining the new unknown vector
\[
  \psi_n(\vep_k)\coloneqq\sqrt{J_n}\,\phi_n(\vep_k)
\]
in terms of
which~\eqref{eq:recurrence:phi} becomes
\begin{equation}
  \label{eq:rrpsi}
  \psi_{n+1}+\frac{\sqrt{J_{n+1}J_{n-1}}}{J_n}\,\psi_{n-1}=2\sqrt{\frac{J_{n+1}}{J_n}}\,\xi_n\psi_n
\end{equation}
(dropping, for simplicity, the $\vep_k$ label for the time being). The continuum limit of this
equation is obtained by setting
\[
  \psi_n\to\psi(x),\quad \psi_{n\pm1}\to\psi(x\pm a),
\]
and similarly for $\xi_n$, $J_n$, $J_{n\pm1}$:
\begin{equation}\label{eq:psixpa}
  \psi(x+a)+\frac{\sqrt{J(x+a)J(x-a)}}{J(x)}\,\psi(x-a)=2\sqrt{\frac{J(x+a)}{J(x)}}\,\xi(x)\psi(x).
\end{equation}
Noting that
\begin{align*}
  \frac{\sqrt{J(x+a)J(x-a)}}{J(x)}&=
                                    \PTqty[\PTqty(1+a\frac{J'(x)}{J(x)}+O(a^2))\PTqty(1-a\frac{J'(x)}{J(x)}+O(a^2))]^{1/2}
                                    =1+O(a^2),\\
  \sqrt{\frac{J(x+a)}{J(x)}}&=\PTqty(1+a\frac{J'(x)}{J(x)}+O(a^2))^{1/2}
                              =1+a\frac{J'(x)}{2J(x)}+O(a^2),
\end{align*}
where the prime denotes derivatives with respect to the spatial coordinate $x$, we can rewrite
eq.~\eqref{eq:psixpa} as
\begin{equation}
  \label{eq:recurrence:psi}
  \psi(x+a)+\psi(x-a)=2\PTqty(1+a\frac{J'(x)}{2J(x)})\xi(x)\psi(x)+O(a^2).
\end{equation}

Motivated by the existence of an arbitrarily small length scale---the lattice spacing $a$---and
the highly oscillatory character of typical solutions, we propose the following WKB-type
ansatz~\cite{DM67} for the solutions of the previous equation:
\begin{equation}
  \psi(x)=\e^{\frac{\iu}{a}S(x)},\qquad
  \text{with}\quad S(x):=\sum_{n=0}^\infty a^nS_n(x)\,.
  \label{eq:ansatz}
\end{equation}
Taylor expanding the exponent
\begin{equation*}
  S_n(x\pm a)=\sum_{m=0}^\infty \frac{(\pm a)^m}{m!}\,S^{(m)}_n(x)
\end{equation*}
and substituting into the first eq.~\eqref{eq:ansatz} we arrive at
\begin{equation}\label{eq:psixpma}
  \psi(x\pm a)=\exp\PTqty[
  \frac{\iu}{a}S_0(x)+\iu\left(S_1(x)\pm 
    S'_0(x)\right)+\iu a\left(\frac{1}{2}S_0''(x)\pm S_1'(x)+S_2(x)\right)+O(a^2)].
\end{equation}
Inserting these expressions into eq.~\eqref{eq:recurrence:psi} and canceling terms we obtain
\begin{equation}
  \label{eq:rrpsifinal}
  \exp\PTqty(\frac{\iu a}{2}S_0''(x))\cos\PTqty(S_0'(x)+a S_1'(x))
  =
  \PTqty(1+a\frac{J'(x)}{2J(x)})\xi(x)+O(a^2).
\end{equation}

We next perform an order by order analysis of the previous equation. At zeroth and first orders we
find, respectively,
\begin{subequations}
  \begin{align}
    \cos\left(S_0'(x)\right)&=\xi(x),\\
    \frac{\iu}2 S_0''(x)\cos\PTqty(S_0'(x))-\sin\PTqty(S_0'(x))S_1'(x)&=
                                                                        \frac{J'(x)}{2J(x)}\xi(x)\,.
  \end{align}
\end{subequations}
These equations have the solution
\begin{subequations}
  \begin{align}
    \label{eq:action_solution:S_0}
    S_0(x,\vep)&=\pm\int_0^x\dd s\arccos\xi(s,\vep)+s_0(\vep),\\
    S_1(x,\vep)&=-\frac{1}{4\iu}\log\left|1-\xi(x,\vep)^2\right|\mp\frac{1}{2}
                 \int_0^x\dd s\frac{\xi(s,\vep)}{\sqrt{1-\xi(s,\vep)^2}}
                 \frac{J'(s)}{J(s)}+s_1(\vep),
  \end{align}
  \label{eq:action_solution}%
\end{subequations}%
where we have restored the dependency on the energy eigenvalue (represented by the continuous
variable $\vep$ in the thermodynamic limit). Dropping the irrelevant overall phase
$s_0(\vep)/a+s_1(\vep)$, our approximate wave functions thus read
\begin{subequations}\label{eq:wkb:general_solution}
  \begin{equation}\label{eq:wkb:exponent}
    \phi(x,\vep)=\frac{A(\vep)}{\sqrt{J(x)
        \PTqty\big|1-\xi(x,\vep)^2|^{1/2}}}\,\e^{\iu\vp(x,\vep)}\,,
  \end{equation}
  where $A(\vep)$ is a normalization constant (which can be taken as positive without loss of
  generality) and
  \begin{equation}
    \vp(x,\vep)=\int_0^x\frac{\dd s}{a}\arccos\xi(s,\vep)+O(a^0)\,.
  \end{equation}
\end{subequations}

In fact, the $\arccos$ function is multivalued, i.e.,
\[
  \arccos z=\pm\Arccos z+2m\pi,\qquad m\in\ZZ,
\]
where $\Arccos$ denotes a particular determination of the inverse cosine. For real
$z$, we shall take
\begin{equation*}
  \Arccos z=
  \begin{cases}
    \pi+\iu\arccosh\abs{z},\quad& z<-1\,,\\
    \arccos z,& \abs{z}\leq 1\,,\\
    \iu\arccosh z,& z>1,
  \end{cases}
\end{equation*}
where in the second line $\arccos:[-1,1]\to[0,\pi]$ is the standard (real) determination. (For
$x<-1$, this determination differs from the principal one~\cite{OLBC10} in the sign of the
imaginary part.) Separating the real and imaginary parts of $\Arccos$ we have
\[
  \Arccos\xi(x,\vep)=\arccos\xi^*(x,\vep)+\iu\,\Th\!\br(\xi(x,\vep)^2-1)\arccosh\abs{\xi(x,\vep)},
\]
where $\Th(s)=(1+\sgn s)/2$ is the Heaviside function and
\begin{equation}
  \xi^*(x,\vep)\coloneqq
  \begin{cases}
    -1, & \vep\leq B(x)-2J(x)\,,\\
    \xi(x,\vep)\equiv\dfrac{\vep-B(x)}{2J(x)},\quad &B(x)-2J(x)<\vep<B(x)+2J(x)\,,\\
    1, & \vep\geq B(x)+2J(x)\,.
  \end{cases}
  \label{eq:xi_clamped}
\end{equation}
Taking $m=0$ we can therefore express our approximate wave function as
\begin{subequations}\label{eq:phi:complex}
\begin{equation}\label{eq:phi:mu}
  \phi(x,\vep)=
  \frac{A(\vep)\,\e^{\pm\mu(x,\vep)}}{\sqrt{J(x)\br|1-\xi(x,\vep)^2|^{1/2}}}\,\e^{\pm\iu\vp^*(x,\vep)},
\end{equation}
where the $\pm$ signs in the exponents are correlated, the phase $\vp^*$ is given by
\begin{equation}
  \vp^*(x,\vep)\coloneqq\int_{0}^x\frac{\dd s}{a}\arccos\xi^*(s,\vep)+O(a^0)\,,
  \label{eq:phase}
\end{equation}
and
\[
  \mu(x,\vep)=\int_0^x\frac{\dd
    s}a\,\Theta\PTqty\big(\xi(s,\vep)^2-1)\,\arccosh\abs{\xi(s,\vep)}\,
  +O(a^0)
\]
\end{subequations}
is real and non-negative. It is important, however, to observe that $\phi$ is defined up to the
overall, position-dependent phase
\begin{equation}\label{eq:theta}
  \th_m(x)=\frac{2m\pi x}a,\qquad m\in\ZZ.
\end{equation}
This ambiguity is unavoidable, since this phase is trivial at the chain's sites $x=na$ (with
$n=0,1,\dots,N-1$), and it will be taken into account in the sequel.

As expected for WKB wave functions, our approximate solution~\eqref{eq:phi:mu} diverges at the
\emph{turning points} determined by the condition $\xi(x,\vep)=\pm1$, or equivalently
\[
  \vep=B(x)\pm 2J(x)\,.
\]
We then have disconnected WKB solutions in each oscillatory/exponential region, respectively
determined by the conditions $\br|\xi(x,\vep)|<1$ or $\br|\xi(x,\vep)|>1$, i.e.,
\[
  B(x)-2J(x)< \vep <B(x)+2J(x)\qquad\text{or}\quad
  \br|\vep-B(x)|>2J(x).
\]
The problem is therefore that of finding appropriate connection formulas for these different
regions (the analogues of potential wells/barriers in the standard WKB solution method).

More precisely, suppose that for a given value of $\vep$ the oscillatory region consists of
$g(\vep)$ disjoint intervals (henceforth referred to as \emph{wells}) $I_i(\vep)$, with
$i=1,\dots,g(\vep)$. Since $\mu$ is constant inside each of these intervals, and the solution
of~\eqref{eq:recurrence:phi} with a real initial condition $\phi_0$ is necessarily real, we can
take
\begin{equation}\label{eq:phii}
  \phi(x,\vep)=\frac{A_i(\vep)\,\sin\vp^*(x,\vep)}{\sqrt{J(x)\br(1-\xi(x,\vep)^2)^{1/2}}}\,,
  \qquad x\in I_i(\vep),
\end{equation}
with $A_i(\vep)$ a real constant. Note that the choice of cosine or sine (or a linear combination
thereof) in the previous formula is purely conventional, since we can absorb an energy dependent
constant phase in the $O(a^0)$ term in the phase $\vp^*(x,\vep)$ (cf. eq.~\eqref{eq:phase}). We
have, however, chosen sine over cosine guided by the behavior of the eigenfunctions in the
homogeneous case (cf.~section~\ref{sec:examples}). At any rate, we shall see in the next section
that this choice is of no consequence for the purposes of approximating the local fermion density,
which depends on $\phi(x,\vep)^2$, due to the highly oscillating nature of this term.

Similarly, if the exponential region $\xi(x,\vep)>1$ consists of a finite number of disjoint
intervals $J_j(\vep)$, inside each of these intervals the solution is given by
\[
  \phi(x,\vep)=\frac{b_j(\vep)\e^{\mu(x,\vep)}+
  c_j(\vep)\e^{-\mu(x,\vep)}}{\sqrt{J(x)\br(\xi(x,\vep)^2-1)^{1/2}}},\qquad x\in J_j(\vep),
\]
with $b_j(\vep)$, $c_j(\vep)$ real. Indeed, in each of these intervals the phase $\vp^*(x,\vep)$
is constant, and can therefore be discarded. Note that, since $\mu(x,\vep)$ is clearly positive by
its very definition, the term $\e^{\mu(x,\vep)}$ is exponentially large and the term
$\e^{-\mu(x,\vep)}$ exponentially vanishing as $a\to0$.

Finally, if the exponential region with $\xi(x,\vep)<-1$ is the disjoint union of a finite number
of intervals $\tilde J_j(\vep)$, inside each of these intervals the real solution can be taken as
\begin{equation}\label{eq:phi:xim1}
  \phi(x,\vep)=\frac{\tilde b_j(\vep)\e^{\mu(x,\vep)}+ \tilde
    c_j(\vep)\e^{-\mu(x,\vep)}}{\sqrt{J(x)\br(\xi(x,\vep)^2-1)^{1/2}}}\,\sin\vp^*(x,\vep),\qquad
  x\in \tilde J_j(\vep),
\end{equation}
with $\tilde b_j(\vep)$, $\tilde c_j(\vep)$ real. Indeed, now the phase $\vp^*$ is \emph{not}
constant on $\tilde J_j(\vep)$, and thus \emph{cannot} be removed. However, as $\pi\to-\pi$ under
$\vp^*\to\vp^*+\th_{-1}$, we can regard the sign of the phase $\pm\vp^*(x,\vep)$ in each of these
intervals as independent of the sign of the exponent $\pm\mu(x,\vep)$, up to an irrelevant overall
constant phase. (Again, the choice of sine or cosine---or a linear combination thereof---in
eq.~\eqref{eq:phi:xim1} is largely a matter of convenience.)

Although connection formulas for the coefficients $A_i$, $b_j$, $c_j$, (or $\tilde b_j$,
$\tilde c_j$) are relatively straightforward to work out~\cite{DM67}, especially if (as is the
case in practice) $g(\vep)$ is small, for our purposes it suffices to note that (much as is the
case in the ordinary WKB method) for $a\to0$ the coefficients $b_j(\vep)$ (or $\tilde b_j(\vep)$)
of the exponentially growing terms in the `forbidden' region $\abs{x,\vep}>1$ become exponentially
small relative to the corresponding coefficients $c_j(\vep)$ (or $\tilde c_j(\vep)$). Thus,
\emph{for all practical purposes the WKB solution~$\phi(x,\vep)$ can be taken as zero in the
  region~$\abs{\xi(x,\vep)}>1$.} As we shall show in the next section, this is, in essence, the
underlying cause of the depletion/saturation phenomenon that we seek to explain
quantitatively.

\begin{remark}
  By the previous observations, as $a\to0$ the wave function should be supported in the region
  $\abs{\xi(x,\vep)}\le1$, i.e., $B(x)-2J(x)\le\vep\le B(x)+2J(x)$. For self-consistency, this
  region should be non-empty for (almost) all single-particle energies $\vep$. Therefore the
  inequalities
  \begin{equation}\label{eq:spec:bounds}
    \min\left(B(x)-2J(x)\right)\le\vep\le\max(B(x)+2J(x))
  \end{equation}
  should provide a heuristic estimate of the range of the single-particle spectrum. In fact,
  Gerschgorin's Circle Theorem~\cite{Ge31} for the eigenvalues of a general (complex) matrix
  applied to $\bH$ yields the \emph{exact} bounds
  \[
    \min\left(B_n-J_n-J_{n-1}\right)\le\vep_k\le\max(B_n+J_n+J_{n-1}),
  \]
  whose continuum limit is eq.~\eqref{eq:spec:bounds}. Furthermore, when $B_n=B$ and $J_n=J$ for
  all $n$ the minimum/maximum single-particle energies are respectively $\vep_0=B-2J$,
  $\vep_{N-1}=B+2J$. This suggests that when $B_n$ and $J_n$ vary slowly we should have
  \begin{equation}\label{eq:eig:bounds}
    \vep_0\underset{N\gg1\vrule width0pt height 6pt}\simeq
    \min\left(B(x)-2J(x)\right),\qquad \vep_{N-1}
    \underset{N\gg1\vrule width0pt height 6pt}\simeq\max(B(x)+2J(x)).
  \end{equation}
  Our numerical simulations do indeed confirm this surmise.\qed
\end{remark}

If we set to zero the wave function in the region~$\abs{\xi(x,\vep)}>1$ (which, as remarked above,
should be an excellent approximation in the limit $a\to0$), the WKB wave function $\phi(x,\vep)$
will be given by the sum of the $g(\vep)$ localized, non-overlapping functions
\begin{equation}\label{eq:phi_i}
  \phi_i(x,\vep)=\frac{A_i(\vep)\sin\vp^*(x,\vep)}{\sqrt{J(x)\br(1-\xi(x,\vep)^2)^{1/2}}}\,
  \chi_i(x)\eqqcolon A_i(\vep) f_i(x,\vep),\qquad i=1,\dots,g(\vep),
\end{equation}
where $\chi_i(x)\coloneqq\chi_{I_i(\vep)}(x)$ is the characteristic function of the interval
$I_i(\vep)$. In practice, unless the `profile' $\xi(x,\vep)$ (which depends on $J_n$, $B_n$, and
$\vep$) is symmetric about the chain's midpoint, or the energy $\vep$ is such that the separation
between consecutive wells is negligible, the eigenfunction of~$\bH$ of energy (close to) $\vep$
will be localized on a single interval roughly coinciding with one of the wells $I_i(\vep)$. Thus
typically the single-particle eigenfunctions will be described by a WKB wave function of the
form~\eqref{eq:phi_i} for a \emph{single} appropriate $i$, which in general depends on the energy
$\vep$. In fact, for many chains of interest---including the Krawtchouk and rainbow chains
discussed in section~\eqref{sec:examples}, and the Rindler chain examined in the Supplementary
Material---the profile $\xi(x,\vep)$ determines only a single well. In this case
eq.~\eqref{eq:phi_i} simplifies to
\begin{subequations}\label{eq:phi:final}
  \begin{equation}
  \label{eq:real:sol}
  \phi(x,\vep)=\frac{A(\vep)\sin\vp^*(x,\vep)}{\sqrt{J(x)\br(1-\xi(x,\vep)^2)^{1/2}}}\,
  \Th\!\br(1-\xi(x,\vep)^2),
\end{equation}
where $A(\vep)>0$ is a normalization constant. As $a\to0$, the normalization constant $A(\vep)$
can be approximated with increasing accuracy by averaging $\sin^2\PTqty(\vp^*(x,\vep))$ over
distances of the order of $a$, with the result
\begin{equation}
  \label{eq:A2vep}
  A(\vep)^{-2}=\int_0^\ell\dd x\,
  \frac{\Th\!\br(1-\xi(x,\vep)^2)}{2J(x)\sqrt{1-\xi(x,\vep)^2}}\,.
\end{equation}
\end{subequations}
We shall see in section~\ref{sec:examples} that the approximation~\eqref{eq:phi:final} yields a
formula for the fermionic density which is in excellent agreement with the numerical results. It
also follows from this equation that the envelopes of the WKB wave function $\phi(x,\vep)$ are in
this case the two curves
\begin{equation}\label{eq:envelope}
  y=\pm\frac{A(\vep)\,\Th\!\br(1-\xi(x,\vep)^2)}{\sqrt{J(x)\br(1-\xi(x,\vep)^2)^{1/2}}}.
\end{equation}
 \begin{remark}
    In the absence of an external magnetic field, the spectrum of the
    chain~\eqref{eq:fermion_Hamiltonian} is symmetric about zero. Moreover, it can be shown (see,
    e.g., \cite{FG21}) that the single-particle wave functions with energies $\vep_k$ and
    $-\vep_k$ are related by
    \begin{equation}\label{eq:phin-vep}
      \phi_n(-\vep_k)=(-1)^n\phi_{n}(\vep_k)
    \end{equation}
    (up to an overall sign). It is straightforward to check that the WKB approximate wave
    functions~\eqref{eq:real:sol} share this symmetry. Indeed, if $B(x)=0$ for all $x$ we have
    $\xi^*(x,-\vep)=-\xi^*(x,\vep)$, and hence
    \[
      a\vp^*(x,-\vep)=\int_0^x\dd s\,\arccos\xi^{*}(x,-\vep)=
      \int_0^x\dd s\,\Bigl(\pi-\arccos\xi^{*}(x,\vep)\Bigr)
      =\pi x-a\vp^*(x,\vep).
    \]
    We can thus write
    \[
      \phi(x,-\vep)=\frac{A(\vep)\sin\br(\frac{\pi
        x}a-\vp^*(x,\vep))}{\sqrt{J(x)\br(1-\xi(x,\vep)^2)^{1/2}}}\,\Th\!\br(1-\xi(x,\vep)^2),
    \]
    since $\xi(x,-\vep)^2=\xi(x,\vep)^2$ when $B=0$ and $A(-\vep)=A(\vep)$ on account of
    eq.~\eqref{eq:A2vep}. In particular, on lattice sites $x=na$ we have
    \begin{equation}\label{eq:phi:PH}
      \phi(na,-\vep)=(-1)^{n+1}\phi(na,\vep),
    \end{equation}
    which is essentially eq.~\eqref{eq:phin-vep}. Obviously, a similar remark can be made for the
    ``localized'' WKB eigenfunctions $\phi_i(x,\vep)$ in eq.~\eqref{eq:phi_i}.\qed
  \end{remark}

\section{Fermionic density profile}\label{sec:spectrum}

As mentioned in section~\ref{sec:prelim}, the single-particle Hamiltonian matrix~$\bH$ and its
corresponding wave functions are related to a finite family of monic orthogonal polynomials
$\{P_n(\vep)\}_{n=0}^N$ (see, e.g., ref.~\cite{FG20}). Indeed, $\bH$ is precisely the Jacobi
matrix associated to this polynomial family, and thus its spectrum consists of the roots of the
polynomial $P_N$ or, in the thermodynamic limit, of the function $\phi(\ell, \vep)$. This
condition is also a direct consequence of the fact that we have imposed Dirichlet boundary
conditions in the recursion relation~\eqref{eq:recurrence:phi}. Therefore in the thermodynamic
limit the single-particle energies should satisfy the condition $\phi(\ell,\vep)=0$. We shall next
use this condition to derive a closed-form asymptotic approximation to the single-fermion density
of states $D(\vep)$.

Let us suppose, to begin with, that the WKB approximation to eigenfunctions with energies in the
range $[\vep,\vep+\dd\vep]$ is supported on a single interval, and is thus given by
eq.~\eqref{eq:real:sol}. If the endpoint $x=\ell$ falls on the region $\abs{\xi(x,\vep)}\le1$, by
eq.~\eqref{eq:phii} we must
have
\begin{equation}
  \label{eq:phiell:vep}
   \vp^*(\ell,\vep)=n\pi,\qquad\text{with}\quad n\in\ZZ.
\end{equation}
It may naively seem that the boundary condition~\eqref{eq:phiell:vep} becomes vacuous when the
endpoint $x=\ell$ falls on a depletion/saturation region $\abs{\xi(x,\vep)}>1$. To see that this
is actually not the case, let us denote the non-depletion interval $\abs{\xi(x,\vep)}<1$ by
$I(\vep)=(x_1(\vep),x_2(\vep))$, with $x_2(\vep)<\ell$. The continuity of the wave function at
$x=x_2(\vep)$ then requires that $\phi(x_2(\vep),\vep)=0$, or equivalently that
\begin{equation}\label{eq:vpx2}
  \vp^*(x_2(\vep),\vep)=k\pi,\qquad\text{with}\quad k\in\ZZ.
\end{equation}
Suppose, first, that $\xi(\ell,\vep)>1$, and hence $\xi>1$ on $[x_2(\vep),\ell]$. Since in this
case the phase $\vp^*$ is constant on the interval $[x_2(\vep),\ell]$, eq.~\eqref{eq:vpx2} implies
\eqref{eq:phiell:vep}. Suppose, next, that $\xi(\ell,\vep)<-1$, and thus $\xi<-1$ on
$[x_2(\vep),\ell]$. Since now $\xi^*=\pi$ on the
interval~$[x_2(\vep),\ell]$, by eq.~\eqref{eq:phase} we have
\[
  \vp^*(x,\vep)=\vp^*(x_2(\vep),\vep)+\frac{\pi}a\,(x-x_2(\vep))
  =k\pi+\frac{\pi}a\,(x-x_2(\vep)),\qquad x_2(\vep)\le x\le\ell,
\]
with $k\in\ZZ$. However, we can change the sign of the last term in the previous equation by
adding $\th_{-1}$ to $\vp^*$, with the result
\[
  \vp^*(x,\vep)=k\pi-\frac{\pi}a\,(x-x_2(\vep)),\qquad x_2(\vep)\le x\le\ell.
\]
Since the phase must be unambiguous at the lattice site $x=\ell$, we must have
\[
  \frac{2\pi}a\,(\ell-x_2(\vep))=2m\pi
\]
for some integer $m$. Hence
\[
  \vp^*(\ell,\vep)=(k+m)\pi,
\]
which is again eq.~\eqref{eq:phiell:vep}.

By eq.~\eqref{eq:phiell:vep}, any two contiguous energy levels must satisfy
\[
  \PTqty\big|\vp^*(\ell,\vep_{k+1})-\vp^*(\ell,\vep_k)|=\pi.
\]
Since the single-particle energy spectrum becomes approximately continuous as $a\rightarrow0$, we
can define an arbitrarily small level spacing $\Delta\vep_k\coloneqq\vep_{k+1}-\vep_k$. From the
previous equation we then deduce that
\begin{equation*}
  \pi=\PTqty\big|\partial_\vep\vp^*(\ell,\vep_k)|\Delta\vep_k+O\br((\Delta\vep_k)^2)\,.
\end{equation*}
On the other hand, from eq.~\eqref{eq:phase} for the phase $\vp^*$ we obtain
\begin{equation}
  \label{eq:pdvepvp}
  \pd_\vep\vp^*(\ell,\vep)=\int_0^\ell\frac{\dd x}a\,\pd_\vep\arccos\xi^*(x,\vep)+O(a^0),
\end{equation}
since $\xi^*(x,\vep)$ is continuous by construction. Taking into account that $\xi^*(x,\vep)$ is
constant in the region where $\abs{\xi(x,\vep)}>1$ we then obtain
\[
  \pd_\vep\vp^*(\ell,\vep)=-\int_0^\ell\frac{\dd x}a\,
  \frac{\Th\!\br(1-\xi(x,\vep)^2)}{2J(x)\sqrt{1-\xi(x,\vep)^2}}+O(a^0)
  =-\frac{A(\vep)^{-2}}a+O(a^0),
\]
with $A(\vep)$ defined in eq.~\eqref{eq:A2vep}. Therefore in the continuum limit
$\De\vep_k\to\De(\vep)$, with
\begin{equation}
  \Delta(\vep)=\frac{\pi}{\PTqty\big|\partial_\vep\vp^*(\ell,\vep)|}
  =\pi aA(\vep)^{2}+O(a^2)\,.
  \label{eq:spacing}
\end{equation}
This result immediately yields the single-fermion density of states
\begin{equation}\label{eq:Dvep}
  D(\vep)=\frac1{\De(\vep)}=\frac{A(\vep)^{-2}}{\pi a}
  =\frac{1}{2\pi a}\int_0^\ell\dd x\frac{\Th\!\br(1-\xi(x,\vep)^2)}{J(x)
    \sqrt{1-\xi(x,\vep)^2}}\,,
\end{equation}
where (as we shall do in the sequel) we have discarded the subleading term $O(a^0)$.

Suppose, next, that for energies in the range $[\vep,\vep+\dd\vep]$ the region
$\abs{\xi(x,\vep)}\le1$ consists of $g(\vep)>1$ wells $I_i(\vep)$. Let us also assume that, as is
almost always the case in practice, the eigenfunctions of $\bH$ with energies in this range are
supported within a single well. Let $D_i(\vep)\dd\vep$ denote the number of eigenfunctions with
energies in the range $[\vep,\vep+\dd\vep]$ supported on the well $I_i(\vep)$. By a slight
modification of the previous argument (essentially, redefining $\xi^*(x,\vep)$ so that it is
constant outside the interval $I_i(\vep)$ and still continuous), we easily arrive at the formula
\begin{subequations}\label{eq:Dvepi}
  \begin{equation}\label{eq:Dvepia}
    D_i(\vep)=\frac{A_i(\vep)^{-2}}{\pi a},
  \end{equation}
  with
  \begin{equation}
    \label{eq:Dvepib}
    A_i(\vep)^{-2}=\int_0^\ell\frac{\chi_i(x)\,\dd x}{2J(x)
      \sqrt{1-\xi(x,\vep)^2}} =\int_{I_i(x)}\frac{\dd x}{2J(x)
      \sqrt{1-\xi(x,\vep)^2}}\,.
  \end{equation}
\end{subequations}
Hence in this case the total single-particle density of states is given by the sum
\begin{equation}
  \label{eq:Dvepmany}
  D(\vep)=\sum_{i=1}^{g(\vep)}D_i(\vep)=\frac1{\pi a}\sum_{i=1}^{g(\vep)}A_i(\vep)^{-2}.
\end{equation}
We shall assume that the previous formula is valid in all generality (i.e., also in the rare cases
in which there are eigenfunctions of $\bH$ supported in several disjoint wells). As we shall see
in the sequel, this assumption is fully supported by the accuracy of the approximations to the
fermion density and the filling fraction in all the examples presented in
section~\ref{sec:examples}.

With the help of the general formula~\eqref{eq:Dvepmany} we can now determine an important
property related to energy, namely the mode filling fraction. Indeed, for a given total occupation
$M$ corresponding to a Fermi energy $\vep_F=\vep_{M-1}$, we may calculate the filling fraction
$\nu$ as
\begin{equation*}
  \nu(\vep_F)=\frac{1}{N}\sum_{k=0}^{M-1}1\rightarrow\frac{a}{\ell}\int_{\vep_0}^{\vep_F}\dd\vep\, D(\vep)\,.
\end{equation*}
We thus obtain
\begin{align}
  \nu(\vep_F)&=\frac{1}{2\pi\ell}\int_{\vep_0}^{\vep_F}\dd\vep\,\sum_{i=1}^{g(\vep)}\int_0^\ell
  \frac{\chi_i(x)\,\dd x}{J(x)\sqrt{1-\xi(x,\vep)^2}}
  =\frac{1}{2\pi\ell}\int_{\vep_0}^{\vep_F}\dd\vep\int_0^\ell\dd x
               \frac{\Th\!\br(1-\xi(x,\vep)^2)}{J(x)\sqrt{1-\xi(x,\vep)^2}}\notag\\
  &=\frac{1}{\pi\ell}\int_0^\ell\dd
    x\int_{-1}^{\xi^*(x,\vep_F)} \frac{\dd\xi}{\sqrt{1-\xi^2}},
    \label{eq:nueff}
\end{align}
where we have taken into account that $\xi(x,\vep_0)=-1$ by eq.~\eqref{eq:eig:bounds}. Evaluating
the innermost integral we obtain
\begin{equation}\label{eq:filling}
  \nu(\vep_F)=\frac{1}{\pi\ell}\int_0^\ell\dd x\,\arccos(-\xi^*(x,\vep_F)),
\end{equation}
or equivalently
\[
\nu(\vep_F)=1-\frac{1}{\pi\ell}\int_0^\ell\dd x\,\arccos\xi^*(x,\vep_F)\\
           =\frac{1}{2}+\frac{1}{\pi\ell}\int_0^\ell\dd x\,\arcsin\xi^*(x,\vep_F),
\]
which are sometimes more convenient forms. Inverting any of these expressions, when possible,
yields the Fermi energy as a function of the filling fraction.

\begin{figure}[h]
  \centering
  \includegraphics[width=\textwidth]{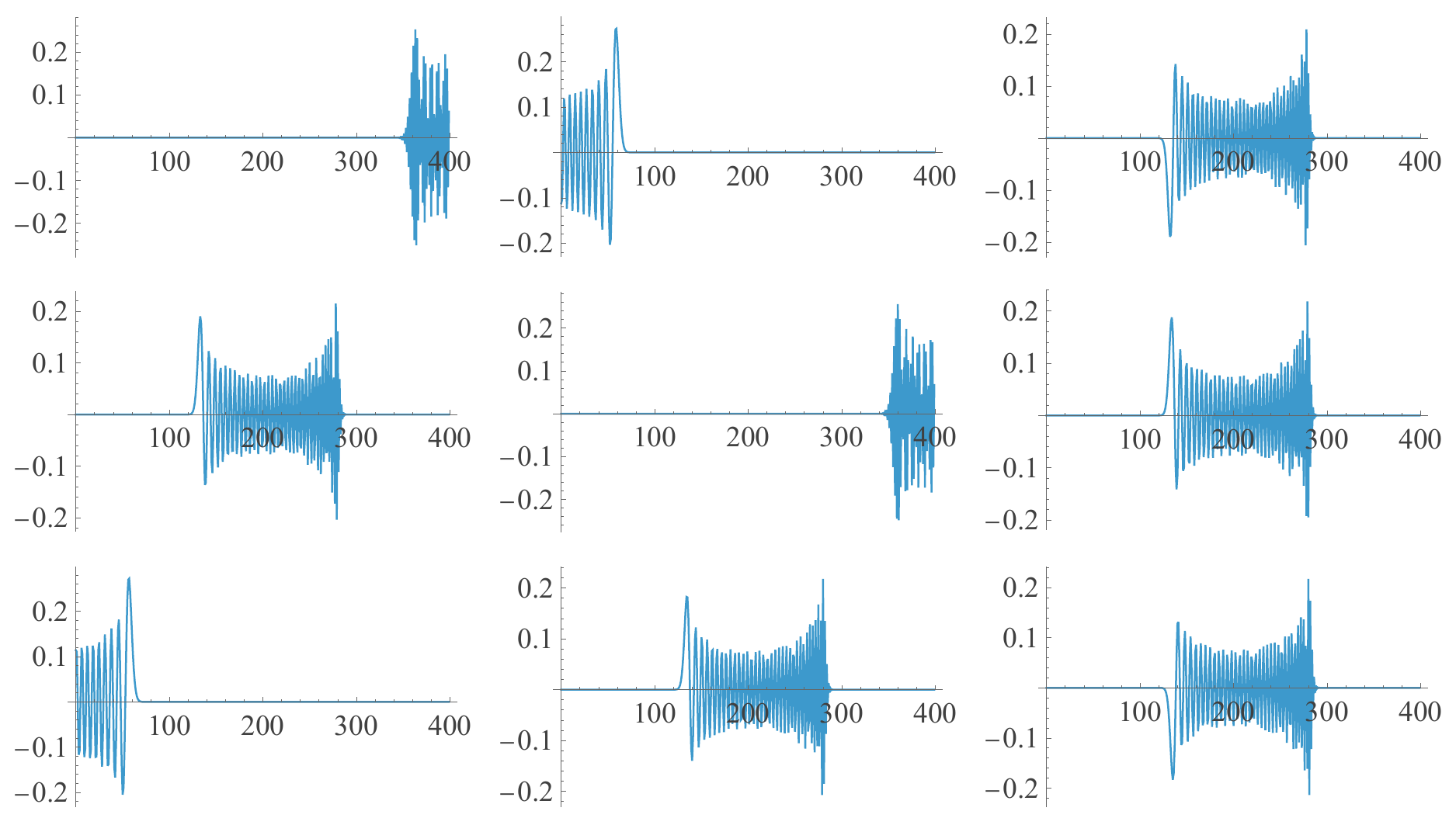}
  \caption{Eigenfunctions of the general cosine chain~\eqref{eq:gen_cos_chain} with $N=400$ spins
    and energies $\vep_{\frac N2+i}$ for $i=-4,-3,\dots,4$ (from top to bottom and left to
  right).}
  \label{fig:frequencies}
\end{figure}

\begin{remark}
  The previous analysis predicts that when the set $\{x\in[0,\ell]:\abs{\xi(x,\vep)}\le1\}$
  consists of $g(\vep)$ disjoint intervals (wells) $I_i(\vep)$, and each single-particle
  eigenfunction with energy near $\vep$ is localized within one of these wells, the relative
  frequency of eigenfunctions localized in the well $I_i(\vep)$ is given by
  $A(\vep)^2/A_i(\vep)^2$. This prediction is fully supported by our numerical results. For
  instance, in fig.~\ref{fig:frequencies} we present a plot of the single-particle eigenfunctions
  of the general cosine chain~\eqref{eq:gen_cos_chain} with $N=400$ spins and energies
  $\vep_{\frac N2+i}$ for $i=-4,-3,\dots,4$, each of which is supported on one of three intervals
  $I_1$ (left end), $I_2$ (middle), and $I_3$ (right end). As can be seen from the figure, the
  number of eigenfunctions supported on each of these wells are respectively $2$, $5$, and $2$.
  The corresponding frequencies $2/9$, $5/9$, and $5/9$ are close to the WKB predictions
  \[
    \frac{A(\vep_{N/2})^2}{A_1(\vep_{N/2})^2}=0.211558,\qquad
    \frac{A(\vep_{N/2})^2}{A_2(\vep_{N/2})^2}=0.547223,\qquad
    \frac{A(\vep_{N/2})^2}{A_3(\vep_{N/2})^2}=0.241218.
  \]
  In fact, if one considers $i=-20,-19,\dots,19$, the corresponding numerical frequencies are
  $0.2$, $0.55$ and $0.25$, in even closer agreement with the previous theoretical values.\qed
\end{remark}

\begin{remark}\label{rem:eff_wf}
  By eq.\eqref{eq:Dvep} and the second equality in eq.\eqref{eq:nueff}, the asymptotic
  formula~\eqref{eq:filling} coincides with the result that would be obtained by assuming that,
  for any energy~$\vep$, eq.~\eqref{eq:real:sol} is the WKB approximation to the exact
  single-particle eigenfunction at that energy. Of course, we know that this is only the case when
  there is only one well for every single-particle energy~$\vep$. In the general case in which the
  region $\abs{\xi(x,\vep)}\le1$ consists of $g(\vep)$ wells, by eq.~\eqref{eq:phi_i} we can
  express the WKB wave function~\eqref{eq:real:sol} as
  \begin{equation}\label{eq:phi-phi_i}
    \phi(x,\vep)=\frac{A(\vep)\,\sin\vp^*(x,\vep)}{\sqrt{J(x)\br(1-\xi(x,\vep)^2)^{1/2}}}
    \sum_{i=1}^{g(\vep)}\chi_i(x)=\sum_{i=1}^{g(\vep)}\frac{A(\vep)}{A_i(\vep)}\,\phi_i(x,\vep).
  \end{equation}
  In other words, in general $\phi(x,\vep)$ is not an approximation to a true eigenfunction, but
  rather a linear combination with weights proportional to $A_i(\vep)^{-1}$ of the single-particle
  approximate WKB eigenfunctions $\phi_i$ supported on each of the wells $I_i(\vep)$. Note also
  that, since both $\phi$ and each $\phi_i$ are normalized, and $\phi_i(x,\vep)$ does not overlap
  with $\phi_j(x,\vep)$ for $j\ne i$, from eq.~\eqref{eq:phi-phi_i} we have
  \[
    A(\vep)^{-2}=\sum_{i=1}^{g(\vep)}A_i(\vep)^{-2}.
  \]
  In particular, when by reasons of symmetry all the normalization constants $A_i(\vep)$ are
  equal, from the previous equations we obtain
  \[
    \phi(x,\vep)=\frac1{\sqrt{g(\vep)}}\sum_{i=1}^{g(\vep)}\phi_i(x,\vep).\eqno\qed
  \]
\end{remark}


We next derive an asymptotic approximation to the fermionic density (per unit length)
\[
\rho(x,\vep_F)\coloneq\frac1a\mel{M}{c^\dagger_{x/a}c^{}_{x/a}}{M},
\]
where $\ket M$ is the state~\eqref{eq:ground_state} filled with single-particle modes with
energies $\vep\le\vep_F$. We can approximate $\rho(x,\vep_F)$ by replacing the sum in
eq.~\eqref{eq:Cnn} by an integral, with the result\footnote{We have again taken into account that
  $\Phi_{nk}\mapsto \sqrt a\,\phi(na,\vep_k)$ as $N\to\infty$.}
\begin{equation}
  \rho(x,\vep_F)=\int_{\vep_0}^{\vep_F}\dd\vep\,\sum_{i=1}^{g(\vep)}D_i(\vep)\phi_i(x,\vep)^2
  =\frac1{\pi a}\int_{\vep_0}^{\vep_F}\dd\vep\,\sum_{i=1}^{g(\vep)}\frac{\phi_i(x,\vep)^2}{A_i(\vep)^2}\,,
  \label{eq:continuum_correlation:simple}
\end{equation}
where we have used eq.~\eqref{eq:Dvepi}. From eq.~\eqref{eq:phi_i} we then obtain
\begin{align*}
  \rho(x,\vep_F)
  &=\frac1{\pi a}\int_{\vep_0}^{\vep_F}\dd\vep\,
    \frac{\sin^2\!\vp^*(x,\vep)}{J(x)\sqrt{1-\xi(x,\vep)^2}}\sum_{i=1}^{g(\vep)}\chi_i(x)\\
  &=\frac1{\pi a}\int_{\vep_0}^{\vep_F}\dd\vep\,
    \frac{\sin^2\!\vp^*(x,\vep)}{J(x)\sqrt{1-\xi(x,\vep)^2}}\,
    \Th\!\br(1-\xi(x,\vep)^2).
\end{align*}
Averaging $\sin^2\!\vp^*(x,\vep)$ over a distance of order $a$ we finally obtain the asymptotic
formula
  \[
    \rho(x,\vep_F)=\frac{1}{2\pi a}\int_{\vep_0}^{\vep_F}\dd\vep\,
    \frac{\Th\!\br(1-\xi(x,\vep)^2)}{J(x)\sqrt{1-\xi(x,\vep)^2}}
    =\frac{1}{\pi a}\int_{-1}^{\xi^*(x,\vep_F)}\frac{\dd\xi}{\sqrt{1-\xi^2}}\,,
  \]
  and therefore
\begin{equation}\label{eq:occupation_density}
  \rho(x,\vep_F)=\frac1{\pi a}\arccos(-\xi^*(x,\vep_F)).
\end{equation}
This approximate expression for the fermionic density profile---together with its discrete
version~\eqref{eq:occ_dens} anticipated in the Introduction---is one of the main results in this
work. Equivalent forms of it that we shall sometimes use are
\[
a\rho(x,\vep_F)
  =1-\frac1{\pi}\arccos\xi^*(x,\vep_F)=\frac{1}{2}
  +\frac1{\pi}\arcsin\xi^*(x,\vep_F).
\]

From eq.~\eqref{eq:occupation_density} and the definition~\eqref{eq:xi_clamped} of $\xi^*$ it
immediately follows that the local fermion density should vanish on the set
$\{x\in[0,\ell]:\vep_F\le B(x)-2J(x)\}$, while it should saturate to one on the set
$ \{x\in[0,\ell]:\vep_F\ge B(x)+2J(x)\}$. The corresponding domains (in general, unions of
disjoint intervals) are thus identified as the \emph{depletion} and \emph{saturation} regions of
the fermionic density, respectively. It is also worth remarking that
eq.~\eqref{eq:occupation_density} is fully consistent with eq.~\eqref{eq:filling}, since the mode
filling $\nu$ can be calculated as
\begin{equation*}
  \nu(\vep_F)=\frac{1}{N}\int_0^\ell\dd x\,\rho(x,\vep_F)
  =\frac{1}{\ell}\int_0^\ell\dd x\,a\rho(x,\vep_F).
\end{equation*}
Note, finally, that at low filling ($\xi^*(x,\vep_F)\gtrsim-1$) we have
\begin{equation}
  \rho(x,\vep_F)=\frac{\sqrt{2}}{\pi a}\sqrt{1+\xi^*(x,\vep_F)}\,
  \bigg[1+O\big(1+\xi^*(x,\vep_F)\big)\bigg]
  \simeq\frac1{\pi a}\sqrt{2+\frac{\vep_F-B(x)}{J(x)}}\,.
\end{equation}
In particular, when the magnetic field is identically zero the previous expression reduces to
eq.~(38) in ref.~\cite{MSSSR22}. It should be stressed, however, that
eq.~\eqref{eq:occupation_density} is valid also in the presence of an external magnetic field and,
perhaps more importantly, for arbitrary fillings.

\begin{remark}
  A popular alternative choice of the free fermionic Hamiltonian~\eqref{eq:fermion_Hamiltonian} is
  \begin{equation}\label{eq:hatH}
    \hat H = -\sum_{n=0}^{N-2}\hat J_n(c^\dagger_nc^{}_{n+1}+\hc)+\sum_{n=0}^{N-1}B_nc^\dagger_n
    c^{}_n,
  \end{equation}
  with $\hat J_n>0$ for all $n$; see, e.g., ref.~\cite{MSSSR22}. (Obviously, this is equivalent to
  taking $J_n<0$ for all $n$ in eq.~\eqref{eq:fermion_Hamiltonian}.) It is straightforward to show
  that if $\phi_n(\vep)$ is a single-particle eigenfunction of $H$ with energy $\vep$ then
  $\hat\phi_n(\vep)=(-1)^n\phi_n(\vep)$ is an eigenfunction of $\hat H$ with the same energy.
  Hence $H$ and $\hat H$ have the same single-particle spectrum, and therefore the filling
  fraction (as a function of the Fermi energy) is identical for both chains. The same is obviously
  true for the average local occupation $\expval*{c^\dagger_nc_n}$, since
  \[
    \expval*{c^\dagger_nc_n}=\sum_{k=0}^{M-1}\phi_n(\vep_k)^2
  \]
  and $\hat\phi_n^2=\phi_n^2$. Thus the Hamiltonians in eqs.~\eqref{eq:fermion_Hamiltonian} and
  \eqref{eq:hatH} have the same fermionic density profile and, in particular, share the same
  depletion/saturation regions. As a consequence, the WKB approximations to the filling fraction,
  the envelopes of the single-particle wave functions, and the fermionic density obtained in this
  and the previous section are valid for the fermionic chain~\eqref{eq:hatH} as well, provided
  that we replace $J(x)$ by $\hat J(x)$.\qed
\end{remark}

\begin{remark}\label{rem:part-hole}%
  When $B(x)$ vanishes identically we obviously have $\xi(x,-\vep)=-\xi(x,\vep)$, and therefore
  $\xi(x,-\vep)\gtrless\pm1$ if and only if $\xi(x,\vep)\lessgtr\mp1$. It follows that
  \[
    \arccos\br(-\xi^*(x,-\vep))=\pi-\arccos\br(-\xi^*(x,\vep)),
  \]
  and hence
  \begin{equation}\label{eq:symm:rhonu}
    a\rho(x,-\vep_F)=1-a\rho(x,\vep_F),\qquad \nu(-\vep_F)=1-\nu(\vep_F).
  \end{equation}
  Thus our approximate formula~\eqref{eq:occupation_density} for the fermionic density is
  consistent with the particle-hole symmetry that occurs when $B(x)$ is identically zero. Of
  course, this property could have also been established directly from eq.~\eqref{eq:phi:PH} for
  the WKB wave functions. If, in addition, $\vep_F=0$ (i.e., at half filling) it follows from the
  previous equation that $a\rho(x,0)=1/2$. We thus recover the well-known result that the
  fermionic density is constant at half-filling and zero magnetic field.\qed
\end{remark}
  \begin{figure}[t!]
  \centering
  \begin{tikzpicture}[xscale=1.6,yscale=.8]
    \draw[-latex] (0,0) -- (3.3,0) node[right] {$x$};
    \draw[very thick,blue] (0,0) -- (3,0) node[below] {\color{black}$\ell$};
    \draw[-latex] (0,-2.8) -- (0,2.8);
    \draw[very thick,red,domain=0:3,samples=400] plot
    (\x,{sin(2*deg(\x))+\x});
    \draw[dashed] (0,1.22837) node[left] {$\vep_1$} -- (2.0944,1.22837);
    \draw[dashed] (0,1.91322) node[left] {$\vep_2$} -- (2.6938,1.91322);
    \draw[dashed] (3,0) -- (3,2.72058);
    \draw[very thick,red] (.2,-1.2) -- (.5,-1.2) node[right]
    {\small\color{black}$B(x)+2J(x)=4J(x)$};
    \draw[very thick,blue] (.2,-1.8) -- (.5,-1.8) node[right]
    {\small\color{black}$B(x)-2J(x)=0$};
  \end{tikzpicture}\hfill
  \begin{tikzpicture}[xscale=1.6,yscale=.8]
    \draw[-latex] (0,0) -- (3.3,0) node[right] {$x$};
    \draw[very thick,red] (0,0) -- (3,0) node[above] {\color{black}$\ell$};
    \draw[-latex] (0,-2.8) -- (0,2.8);
    \draw[very thick,blue,domain=0:3,samples=400] plot
    (\x,{-sin(2*deg(\x))-\x});
    \draw[dashed] (0,-1.22837) node[left] {$\vep_2$} -- (2.0944,-1.22837);
    \draw[dashed] (0,-1.91322) node[left] {$\vep_1$} -- (2.6938,-1.91322);
    \draw[dashed] (3,0) -- (3,-2.72058);
    \draw[very thick,red] (.2,1.8) -- (.5,1.8) node[right]
    {\small\color{black}$B(x)+2J(x)=0$};
    \draw[very thick,blue] (.2,1.2) -- (.5,1.2) node[right]
    {\small\color{black}$B(x)-2J(x)=-4J(x)$};
  \end{tikzpicture}
  
  \caption{Plots of $B(x)\pm 2J(x)$ for $B(x)=2J(x)$ (left) and $B(x)=-2J(x)$ (right), for a
    continuum coupling $J(x)$ vanishing at the origin. In the former case there is a single
    saturation interval $[0,x_1(\vep_F)]$ for Fermi energies $\vep_F\in[0,\vep_1]$, where
    $x_1(\vep_F)$ is the smallest positive root of the equation $4J(x)=\vep_F$. A second, disjoint
    saturation interval $[x_2(\vep_F),x_3(\vep_F)]$ appears when $\vep_F\in(\vep_1,\vep_2)$, where
    $x_2(\vep_F)$ and $x_3(\vep_F)$ are the second and third smallest positive roots of the
    equation $4J(x)=\vep_F$ . These two saturation regions merge into a single interval
    $[0,x_1(\vep_F)]$ for energies above $\vep_2$. The situation is similar for the case
    $B(x)=-2J(x)$ represented in the right panel, the saturation intervals becoming depletion
    ones.}
  \label{fig:sat:dep}
\end{figure}
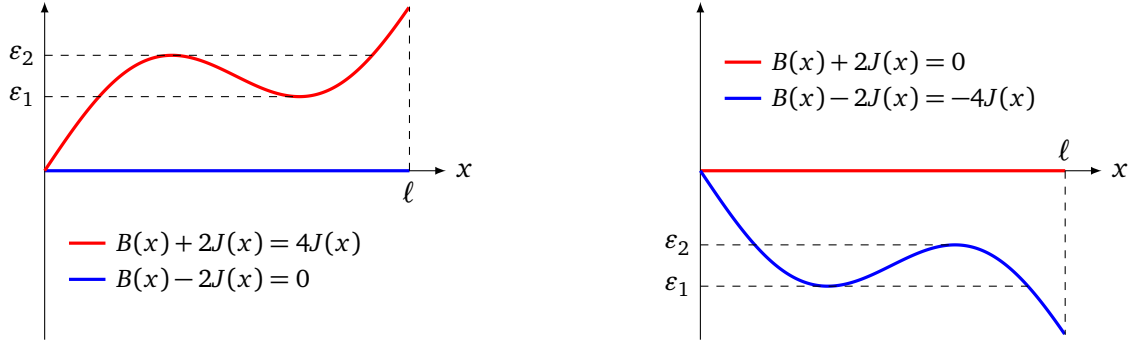%
  Equation~\eqref{eq:occupation_density} allows us to deduce interesting global properties of the
  fermionic density that are difficult to predict a priori, in particular regarding the influence
  of the magnetic field on the depletion/saturation phenomenon. For example, in the absence of an
  external magnetic field the depletion and saturation regions are determined by the equations
  $\vep_F\le -2J(x)$ and $\vep_F\ge 2J(x)$, respectively. Thus there are no depletion and
  saturation intervals for Fermi energies
  $\vep_F\in[-2\min\limits_{x\in[0,\ell]}J(x),2\min\limits_{x\in[0,\ell]}J(x)]$. Moreover, for
  $\vep_F\in[-2\max_{x\in[0,\ell]}J(x),-2\min\limits_{x\in[0,\ell]}J(x)]$ only depletion regions
  can appear, while for $\vep_F\in[2\min\limits_{x\in[0,\ell]}J(x),2\max_{x\in[0,\ell]}J(x)]$ only
  saturation intervals may occur.

  In the same vein, using eq.~\eqref{eq:occupation_density} it is a straightforward matter to
  construct chains without depletion or saturation regions for \emph{all} fillings. Indeed, if
  $J(0)=0$ and $B(x)=2J(x)$ by eq.~\eqref{eq:occupation_density} there are no depletion regions
  and there is at least one saturation region at the left end of the chain (see
  fig.~\ref{fig:sat:dep} left). Moreover, if $J$ is monotonically increasing there are no other
  saturation regions; this is the case, for instance, with the Rindler chain discussed in the
  Supplementary Material. Similarly, taking $B(x)=-2J(x)$ we obtain a chain with no saturation
  regions and at least one depletion region at the right end of the chain, with no other such
  regions if $J$ is monotonic (cf.~fig.~\ref{fig:sat:dep}, right panel). (Of course, to construct
  chains with only depletion or saturation intervals respectively growing from the left or right
  endpoint, it suffices to take $J(\ell)=0$ instead of $J(0)=0$.)

\section{Examples}\label{sec:examples}

\subsection{The homogeneous chain}

Consider, to begin with, the homogeneous chain with
\[
  B_n=B,\qquad J_n=J
\]
for all $n$. As mentioned in section~\ref{sec:WKB}, in this case the single-particle spectrum is
known in closed form~\cite{EP18}. More precisely, the single-particle energies (in increasing
order) are given by
\begin{equation}\label{eq:eigv:hom}
  \vep_{k-1}=B-2J\cos\biggr(\frac{\pi k}{N+1}\biggr),\qquad k=1,\dots,N,
\end{equation}
and their corresponding (normalized) eigenfunctions~$\sum_{n=1}^{N}\Phi_{n-1,k-1}\ket{n-1}$ have
components
\begin{equation}\label{eq:Phink:hom}
  \Phi_{n-1,k-1}=(-1)^{n-1}\sqrt{\frac2{N+1}}\,\sin\biggl(\frac{\pi
    nk}{N+1}\biggr),\qquad n,k=1,\dots,N\,.
\end{equation}
In this case
\[
  \xi(x,\vep)=\frac{\vep-B}{2J}\equiv\xi(\vep)
\]
is independent of $x$, and $\abs{\xi(\vep)}\le1$  on account of eq.~\eqref{eq:eigv:hom}. It follows
that $\xi^\star(x,\vep)=\xi(\vep)$, and therefore
\[
  \vp^*(x,\vep)=\int_0^x\frac{\dd s}a\,\arccos\xi(\vep)=\frac{x}a\arccos\xi(\vep).
\]
Moreover, by eq.~\eqref{eq:A2vep} we have
\[
  A(\vep)^{-2}=\frac{\ell}{2J\sqrt{1-\xi(\vep)^2}},
\]
and hence the approximate WKB eigenfunctions in this case reduce to
\[
  \phi(x,\vep)=\sqrt{\frac2\ell}\,\sin\!\br(\frac{x}a\arccos\xi(\vep)).
\]
On the other hand,
\begin{align*}
  \sqrt{a}\,\phi(na,\vep_{k-1})
  &=\sqrt{\frac2N}\,\sin\!\br(\!n\arccos(\frac{\vep_{k-1}-B}{2J}))
    =\sqrt{\frac2N}\,\sin\!\br(\!n\arccos\br(-\cos\!\br(\frac{\pi k}{N+1})))\\
  &=\sqrt{\frac2N}\,\sin\!\br(\!n\pi-\frac{\pi nk}{N+1})=(-1)^{n-1}\sqrt{\frac2N}\,
    \sin\!\br(\frac{\pi nk}{N+1})\\
  &=\sqrt{\frac{N}{N+1}}\,\Phi_{n-1,k-1}\,,
\end{align*}
so that the relative error in the WKB wave function is $O(N^{-1})$.

By eq.~\eqref{eq:eigv:hom}, the filling $\nu=M/N$ (with $M=1,\dots,N$) is in this case given by
\[
  \nu(\vep_{M-1})=\frac{N+1}{N\pi}\,\arccos(-\xi_{M-1}).
\]
This should be compared with our approximate formula~\eqref{eq:filling}, which in this case
(taking $\vep_F=\vep_{M-1}$) reads
\[
  \nu(\vep_{M-1})=\frac1{\pi\ell}\,\ell\arccos(-\xi_{M-1})=\frac1{\pi}\,\arccos(-\xi_{M-1}),
\]
with a relative error again of order $N^{-1}$. Likewise, the level spacing
$\De\vep_{k-1}=\vep_{k}-\vep_{k-1}$ can be directly computed from eq.~\eqref{eq:eigv:hom}, with
the result
\[
  \De\vep_{k-1}=2J\left[\cos\!\br(\frac{\pi k}{N+1})-\cos\!\br(\frac{\pi
      (k+1)}{N+1})\right]\simeq\frac{2\pi J}{N+1}\sin\!\br(\frac{\pi k}{N+1}).
\]
On the other hand, its continuous approximation~\eqref{eq:spacing} is given by
\[
  \De(\vep)=\pi a A(\vep)^{2}=\frac{2\pi J}{N}\sqrt{1-\xi(\vep)^2},
\]
which evaluated at $\vep=\vep_{k-1}$ yields
\[
  \De(\vep_{k-1})=\frac{2\pi J}{N}\sqrt{1-\left(\frac{\vep_{k-1}-B}{2J}\right)^2}
  =\frac{2\pi J}{N}\sin\!\br(\frac{\pi k}{N+1})=\De\vep_{k-1}+O(N^{-1}).
\]

Consider, finally, the approximation~\eqref{eq:occupation_density} to the fermionic density, which
in this case is simply
\[
  \rho(x,\vep_F)=\frac1{\pi a}\,\arccos\br(-\xi(\vep_F)).
\]
Taking $\vep_{F}=\vep_{M-1}$ for a filling fraction $\nu=M/N$ we obtain the WKB
approximation~\eqref{eq:occ_dens} to the average number of fermions at site $n$:
\[
  \expval*{c^\dagger_nc_n}\equiv\expval{c^\dagger_nc_n}{M}\approx a\rho(na,\vep_{M-1})
  =\frac1\pi\,\arccos\br(-\xi(\vep_{M-1}))
  =\frac1\pi{\frac{\pi M}{N+1}}=\nu+O(N^{-1}).
\]
In other words, in the thermodynamic limit the fermionic density should be constant for any
filling and magnetic field strength. For finite $N$, the fermionic density can be exactly computed
with the help of the explicit formula~\eqref{eq:Phink:hom}, namely
\[
  \expval*{c^\dagger_{n-1}c_{n-1}}=\frac{2}{N+1}\sum_{k=0}^{M-1}\sin^2\br(\frac{\pi nk}{N+1})=
  \frac{M}{N+1}
  -\frac{\sin\!\br(\frac{M\pi n}{N+1})
    \cos\!\br(\frac{(M-1)\pi n}{N+1})}{(N+1)\sin\!\br(\frac{\pi n}{N+1})}\,.
\]
Thus in the bulk (i.e., for $n$ away from $1$ and $N$) and in the thermodynamic limit the average
number of fermions is indeed independent of the site---and, as a consequence, simply equal to the
filling fraction $\nu$. As expected, $\expval*{c^\dagger_{n-1}c_{n-1}}$ features characteristic
Friedel oscillations~\cite{Fr58} near the chain's ends. For example, for small $n\ge1$ we have
\[
  \expval*{c^\dagger_{n-1}c_{n-1}}\approx\frac{M}{N+1}-\frac{\sin\!\br(\frac{2M\pi
      n}{N+1})}{2\pi n}\,,
\]
which oscillates around $M/(N+1)$ with the Friedel frequency $2M\pi/(N+1)$ and amplitude
proportional to $n^{-1}$.

In summary, all our approximate formulas from the previous section are in excellent agreement with
the exact results in the homogeneous case.
\begin{remark}
  The results for the homogeneous XX chain discussed in this example can be also used to provide a
  heuristic justification of eq.~\eqref{eq:occupation_density} in the spirit of the local density
  approximation (LDA)~\cite{BBEPV25}. The results for the homogeneous XX chain discussed in this
  example can be also used to provide a heuristic justification of
  eq.~\eqref{eq:occupation_density} in the spirit of the local density approximation (LDA)
  formalism~\cite{BBEPV25}. Indeed, in the homogeneous case we can rewrite eq.~\eqref{eq:eigv:hom}
  as
  \begin{equation}\label{eq:vep_p}
    \vep_{k-1}=B-2J\cos(a p_{k-1}),\qquad k=1,\dots,N,
  \end{equation}
  where the effective Fermi momentum $p_{k-1}$ is given by
  \[
    p_{k-1}=\frac{\pi k}{a(N+1)}
  \]
  (note that we have introduced the chain spacing $a$ for dimensional reasons). In this case the
  fermionic density (per unit length) for the Fermi energy $\vep_{k-1}$ is constant, given by
  \[
    \rho_{k-1}=\frac{k}{Na}.
  \]
  By eq.~\eqref{eq:vep_p}, in the homogeneous case the relation between the Fermi momentum and the
  Fermi energy is given by
  \begin{equation}\label{eq:pk_vep}
    p_{k-1}=\frac1a\,\arccos\left(\frac{B-\vep_{k-1}}{2J}\right)
  \end{equation}
  Following the LDA approach, we promote $p_{k-1}$ to a local Fermi momentum $p(x,\vep_F)$.
  The local version of eq.~\eqref{eq:pk_vep} then yields the dispersion relation for $p(x,\vep_F)$:
  \[
    p(x,\vep_F)=\frac1a\,\arccos\left(\frac{B(x)-\vep_F}{2J(x)}\right).
  \]
  Similarly, in the homogeneous case the (constant) fermionic density $\rho_{k-1}$ for a Fermi
  energy $\vep_{k-1}$ is related to the Fermi momentum by
  \[
    \rho_{k-1}=\frac{N+1}{N}\frac{p_{k-1}}\pi\simeq \frac{p_{k-1}}\pi.
  \]
  The continuous version of this formula
  \[
    \rho(x,\vep_F)=\frac1\pi\,p(x,\vep_F)=\frac1{\pi a}\,
    \arccos\left(\frac{B(x)-\vep_F}{2J(x)}\right)
  \]
  coincides with eq.~\eqref{eq:occupation_density} away from the saturation/depletion regions
  $\abs{(B(x)-\vep)/(2J(x))}>1$. The behavior of the fermionic density in these regions (or, more
  precisely, whether $\rho(x,\vep_F)$ is zero or one) can be inferred from the previous formula,
  for example, by a continuity argument, which is equivalent to taking the real part of the
  complex $\arccos$ function. Note, however, that by their very nature arguments based on LDA
  cannot provide precise information on other properties of interest, as for instance the behavior
  of the single particle wave functions.\qed
\end{remark}

\subsection{The Krawtchouk chain}\label{subsec:Kr}

We shall next consider the Krawtchouk chain~\cite{CNV19,FG21,BPV24}, whose hoppings and magnetic
field are given by
\begin{equation}\label{eq:JBKraw}
  J_n=\sqrt{q(1-q)(n+1)(N-n-1)},\qquad B_n=(N-1)q+(1-2q)n
\end{equation}
with $0<q<1$. This inhomogeneous chain is also exactly solvable. Indeed, the single-particle
energies are the integers $0,\dots,N-1$, with corresponding eigenfunctions
\begin{equation*}
  \phi_{n}(\vep_k)=(-1)^n\sqrt{\binom{N-1}k\binom{N-1}n}\,q^{\frac12(k+n)}
  (1-q)^{\frac12(N-k-n-1)}\,K_n(k;q,N-1).
    \label{PhiKraw}
\end{equation*}
Here  $K_n(x;q,N-1)$ is the Krawtchouk polynomial defined by~\cite{KLS10}
\begin{equation}\label{KrawP}
  K_n(x;q,N-1)={}_2F_1(-n,-x;1-N;1/q)\,,\qquad n=0,\dots,N-1\,,
\end{equation}
where ${}_2F_1$ is the standard hypergeometric function
\[
  {}_2F_1(a,b;c;z)\coloneqq\sum_{k=0}^\infty\frac{(a)_k(b)_k}{(c)_k}\,\frac{z^k}{k!}\,,
\]
$(a)_k\coloneqq a(a+1)\cdots(a+k-1)$ being the (ascending) Pochhammer symbol. After a rescaling by
$1/N$, the continuum limit of $J_n$ and $B_n$ are given by
\[
  N^{-1}J_n\to J(x)=\sqrt{q(1-q)\frac{x}\ell\br(1-\frac{x}\ell)},\qquad
  N^{-1}B_n\to B(x)=q+(1-2q)\frac{x}\ell\,.
\]
Consequently, the rescaled single-particle spectrum is
\begin{equation}\label{eq:vepKraw}
  \vep_k=\frac{k}N,\qquad k=0,\dots,N-1,
\end{equation}
so that the level spacing
\begin{equation}\label{eq:Kraw:spacing}
  \De\vep_k=\frac1N
\end{equation}
is constant. In particular, with this normalization
\[
  \nu=\frac{M}N=\vep_{M}=\vep_F+\frac1N,
\]
and thus $\nu(\vep_F)=\vep_F$ in the thermodynamic limit.

An elementary calculation that we shall omit shows that in this case
\[
  \min\bigl\{B(x)-2J(x):x\in[0,\ell]\bigr\}=0,\qquad \max\bigl\{B(x)+2J(x):x\in[0,\ell]\bigr\}=1,
\]
\begin{figure}[t!]
\centering
  \begin{tikzpicture}[scale=3.5]
    \newcommand{\q}{.25}
    \draw[-latex] (0,0) -- (1.25,0) node[right] {$x/\ell$};
    \draw[-latex] (0,0) -- (0,1.25) node[above] {$B(x)\pm2J(x)$};
    \draw[very thick,blue,domain=0:1,samples=300] plot
    (\x,{\q+(1-2*\q)*\x-2*sqrt(\q*(1-\q)*\x*(1-\x))});
    \draw[very thick,red,domain=0:.5,samples=300] plot
    (\x,{\q+(1-2*\q)*\x+2*sqrt(\q*(1-\q)*\x*(1-\x))});
    \draw[very thick,red,domain=0:.5,samples=300] plot
    (1-\x,{\q+(1-2*\q)*(1-\x)+2*sqrt(\q*(1-\q)*\x*(1-\x))});
    \draw[thick,dashed] (1,0) node[below] {$1$}-- (1,1);
    \draw[thick,dashed] (0,1-\q) node[left] {$1-q$}-- (1,1-\q);
    \draw[] (-.02,\q) node[left] {$q$} -- (.02,\q);
    \draw[] (\q,-.02) node[below] {$q$} -- (\q,.02);
    \draw[thick,dashed] (1-\q,0) node[below] {$1-q$} -- (1-\q,1);
    \draw[thick,dashed] (0,1) node[left] {$1$} -- (1,1);
  \end{tikzpicture}
  \qquad
    \begin{tikzpicture}[scale=3.5]
    \newcommand{\q}{.75}
    \draw[-latex] (0,0) -- (1.25,0) node[right] {$x/\ell$};
    \draw[-latex] (0,0) -- (0,1.25) node[above] {$B(x)\pm2J(x)$};
    \draw[very thick,blue,domain=0:.5,samples=300] plot
    (\x,{\q+(1-2*\q)*\x-2*sqrt(\q*(1-\q)*\x*(1-\x))});
    \draw[very thick,blue,domain=0:.5,samples=300] plot
    (1-\x,{\q+(1-2*\q)*(1-\x)-2*sqrt(\q*(1-\q)*\x*(1-\x))});
    \draw[very thick,red,domain=0:.5,samples=300] plot
    (\x,{\q+(1-2*\q)*\x+2*sqrt(\q*(1-\q)*\x*(1-\x))});
    \draw[very thick,red,domain=0:.5,samples=300] plot
    (1-\x,{\q+(1-2*\q)*(1-\x)+2*sqrt(\q*(1-\q)*\x*(1-\x))});
    \draw[thick,dashed] (1,0) node[below] {$1$}-- (1,1);
    \draw[thick,dashed] (0,1-\q) node[left] {$1-q$}-- (1,1-\q);
    \draw[] (-.02,\q) node[left] {$q$} -- (.02,\q);
    \draw[] (\q,-.02) node[below] {$q$} -- (\q,.02);
    \draw[thick,dashed] (1-\q,0) node[below] {$1-q$} -- (1-\q,1);
    \draw[thick,dashed] (0,1) node[left] {$1$} -- (1,1);
  \end{tikzpicture}
  \caption{Plot of $B(x)+2J(x)$ and $B(x)-2J(x)$ (respectively in red and blue) for the Krawtchouk
    chain of unit length with $0<q\le1/2$ (left) and $1/2\le q\le 1$ (right).}
  \label{fig:Kraw}
\end{figure}
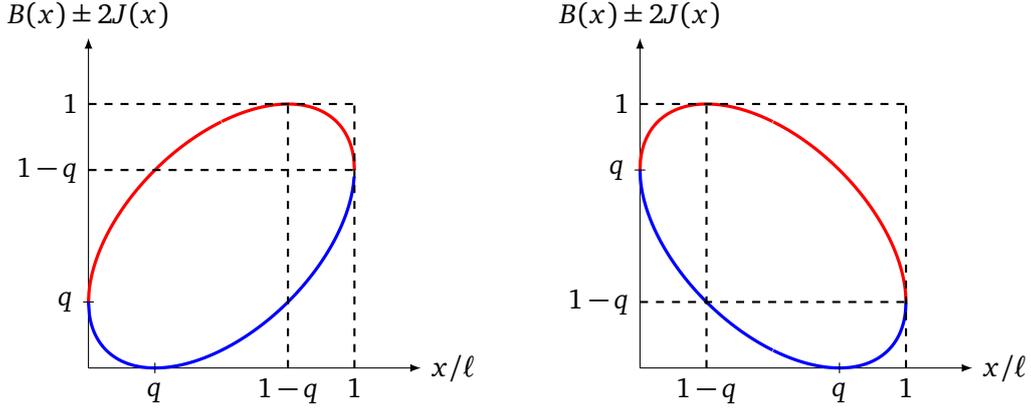%
for all $q\in[0,1]$, as expected since the single-particle spectrum is the interval $[0,1]$. In
fact, the curve $y=B(x)\pm 2J(x)$ is an ellipse inscribed in the rectangle $[0,\ell]\times[0,1]$
and tangent to it at the extremal points $(\ell q,0)$ and $(\ell(1-q),1)$
(cf.~fig.~\ref{fig:Kraw}).

We shall next evaluate the asymptotic approximation~\eqref{eq:spacing} to the level spacing. To
begin with, from fig.~\ref{fig:Kraw} it follows that in this case the region
$\abs{\xi(x,\vep)}\le1$ consists of a single interval $[x_1(\vep),x_2(\vep)]$, where
$x_{1,2}(\vep)$ are the abscissas of the two intersection points of the horizontal line $y=\vep$
with the ellipse $(y-B(x))^2-4J^2(x)=0$. In other words, $x_{1,2}(\vep)$ are the roots of the
second-degree polynomial
\begin{equation}\label{eq:Pxvep}
  P(x,\vep)\coloneqq(\vep-B(x))^2-4J^2(x)=\frac{x^2}{\ell^2}
  -2\br[\vep(1-2q)+q]\frac{x}\ell+(\vep-q)^2.
\end{equation}
It follows that
\[
  (\vep-B(x))^2-4J^2(x)=\frac1{\ell^2}\br\big(x-x_1(\vep))\br\big(x-x_2(\vep)),
\]
and therefore, by eq.~\eqref{eq:A2vep},
\begin{align*}
  A(\vep)^{-2}&=\int_{x_1(\vep)}^{x_2(\vep)}\frac{\dd x}{\sqrt{4J(x)^2-(\vep-B(x)^2)}}
  =\int_{x_1(\vep)}^{x_2(\vep)}\frac{\dd x}{\sqrt{-P(x,\vep)}}\\
  &=\ell\int_{x_1(\vep)}^{x_2(\vep)}\frac{\dd x}{\sqrt{(x-x_1(\vep))(x_2(\vep)-x)}}.
\end{align*}
The integral is easily computed with the standard change of variable
\[
  x=\frac12\br\big(x_1(\vep)+x_2(\vep))+\frac12\br\big(x_2(\vep)-x_1(\vep))\sin t,
\]
and its value is $\pi$ (independent of $\vep$). We thus obtain
\begin{equation}\label{eq:De:Kraw}
  \De(\vep)=\pi aA(\vep)^2=\frac{a}\ell=\frac1N\,.
\end{equation}
Therefore in this case the asymptotic formula~\eqref{eq:spacing} (to lowest order in $a$) gives
the exact value~\eqref{eq:Kraw:spacing}.

\begin{figure}[t!]
  \includegraphics[width=.48\textwidth]{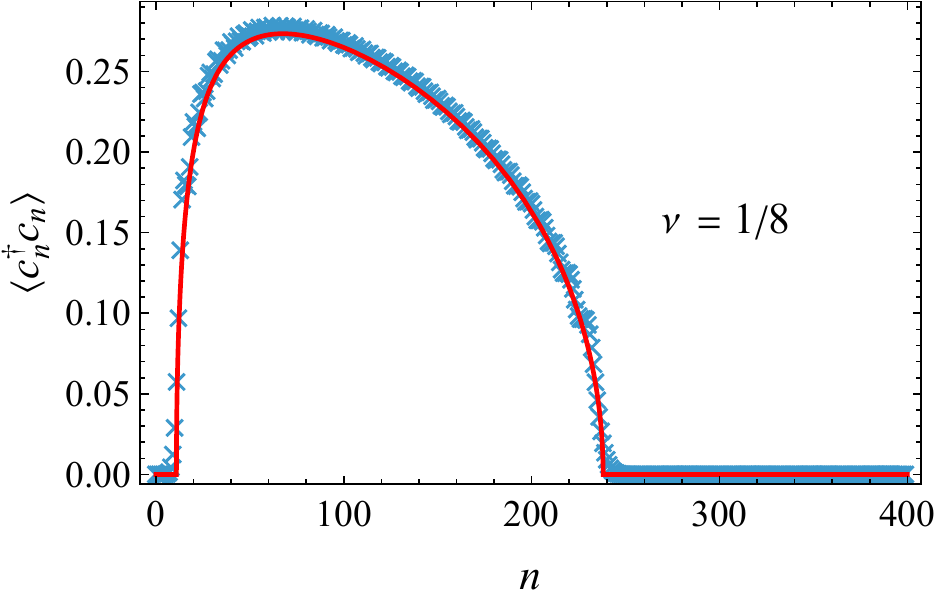}\hfill
  \includegraphics[width=.48\textwidth]{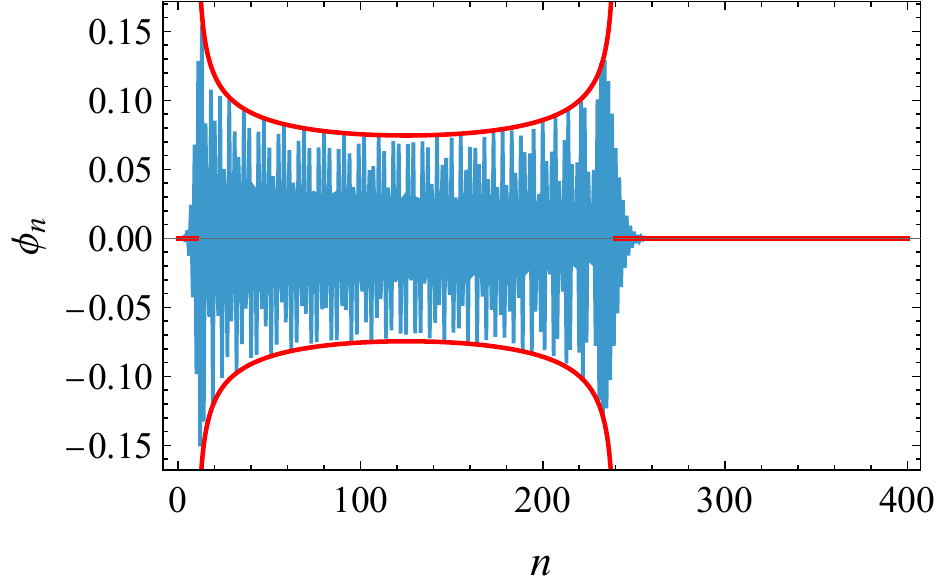}\\
  \includegraphics[width=.48\textwidth]{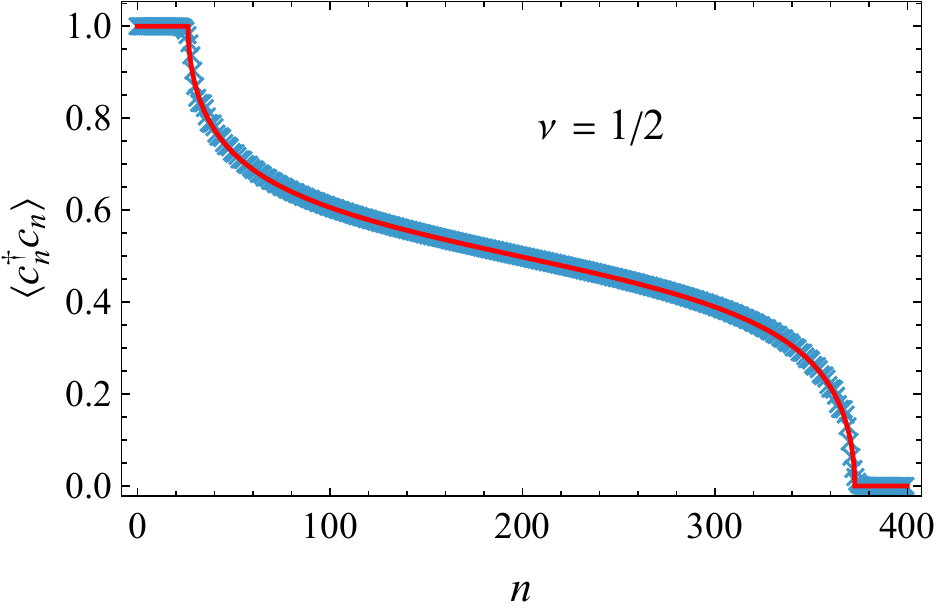}\hfill
  \includegraphics[width=.48\textwidth]{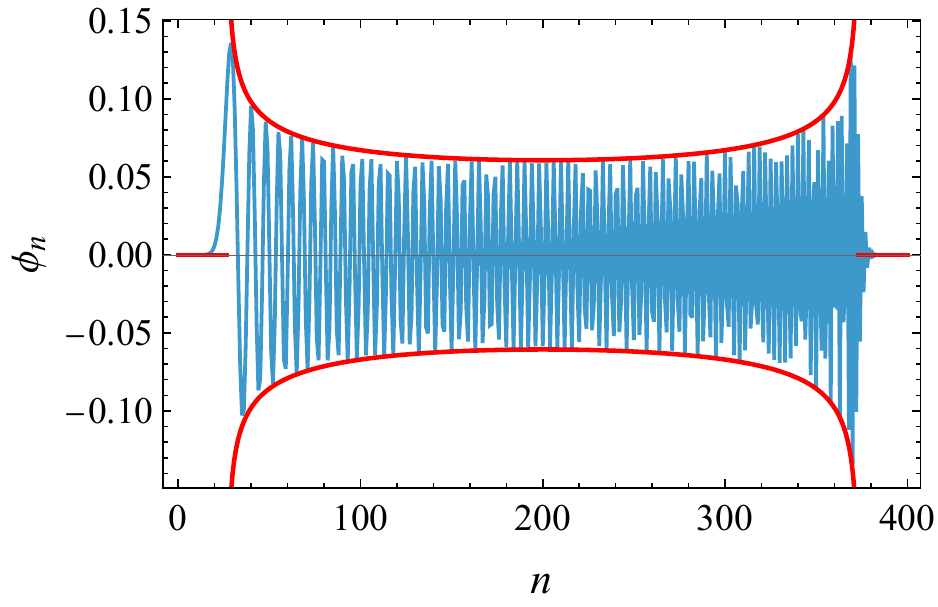}\\
  \includegraphics[width=.48\textwidth]{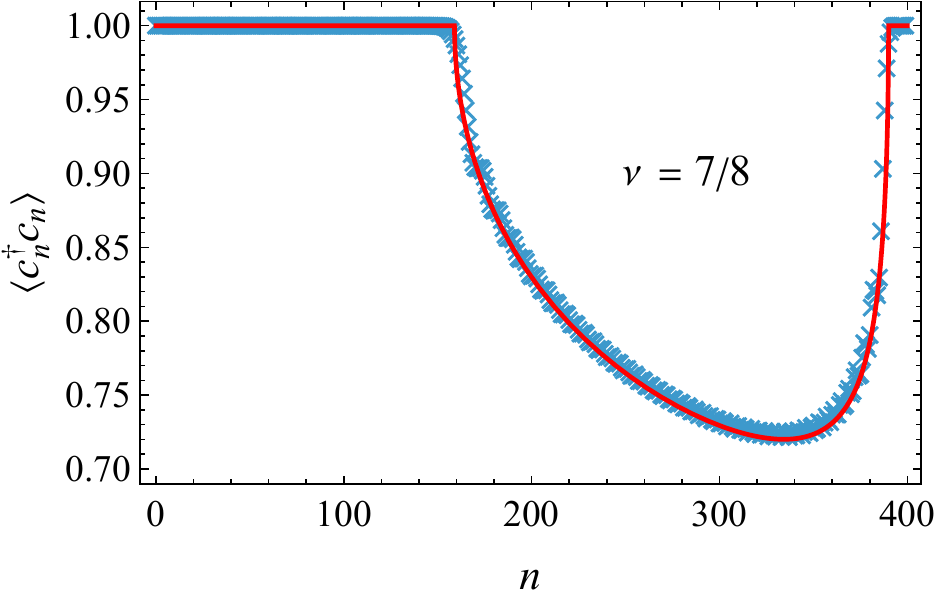}\hfill
  \includegraphics[width=.48\textwidth]{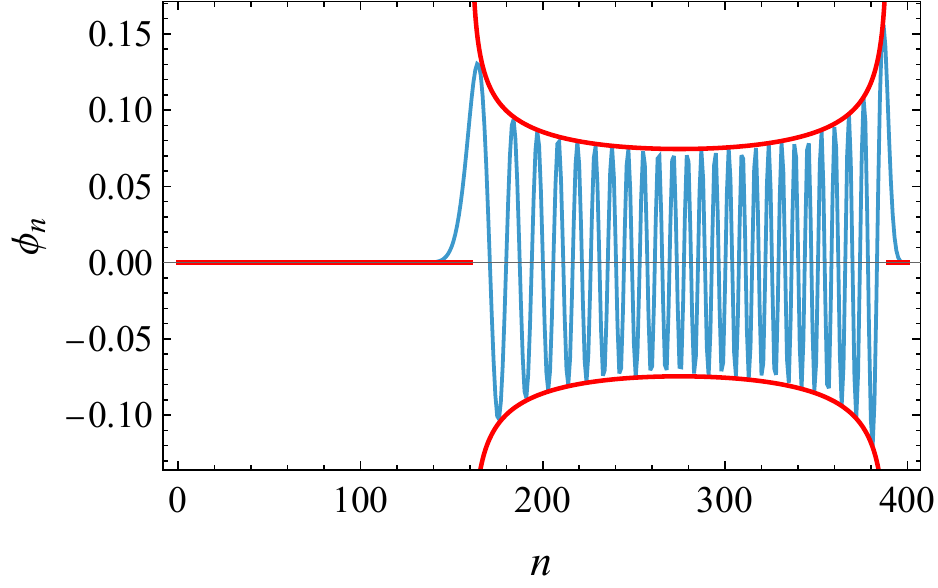}
  \caption{Left column: local fermion density for the Krawtchouk chain with $q=1/4$, $N=400$, and
    fillings $\nu=1/8$, $1/2$, and $7/8$ (blue markers), compared to its asymptotic
    approximation~\eqref{eq:rho:Kraw} (red line). Right column: analogous plot of the
    corresponding single-particle eigenfunctions $\phi_n(N\nu)$ (in blue) compared to their WKB
    envelopes~\eqref{eq:phi:env} (red lines).}
  \label{fig:Kraw:rho}
\end{figure}

From the previous calculation it readily follows that the envelopes~\eqref{eq:envelope} of the WKB
wave functions~\eqref{eq:phi:final} are in this case the curves
\begin{subequations}\label{eq:phi:env}
  \begin{equation}\label{eq:phi:prof}
    y=\pm\sqrt{\frac2\pi}\,\br\big[(x-x_1(\vep))(x_2(\vep)-x)]^{-1/4},
    \qquad x_1(\vep)<x<x_2(\vep),
  \end{equation}
  where the turning points $x_{1,2}(\vep)$ are explicitly given by
  \begin{equation}\label{eq:x12:Kraw}
    \frac{x_{1,2}(\vep)}\ell=q+\vep(1-2q)\pm2\sqrt{q(1-q)\vep(1-\vep)};
  \end{equation}
\end{subequations}
cf.~eq.~\eqref{eq:Pxvep} and fig.~\ref{fig:Kraw:rho} (right column), where (as in the remainder of
this section) we have taken $a=1$, $\ell=N$.

Let us next analyze the behavior of the fermionic density using our asymptotic
formula~\eqref{eq:occupation_density}. Since the discussion is very similar for the cases
$0<q\le1/2$ and $1/2\le q<1$, we shall restrict in what follows to the former case. First of all,
from fig.~\ref{fig:Kraw} it is clear that for $\vep_F\in[0,1]$ we have
\begin{equation}\label{eq:rho:Kraw}
  \rho(x,\vep_F)=\frac1{\pi a}\arccos\br(\frac{B(x)-\vep_F}{2J(x)}),
  \qquad x_1(\vep_F)\le x\le x_2(\vep_F),
\end{equation}
where the turning points $x_{1,2}(\vep_F)$ are given by eq.~\eqref{eq:x12:Kraw}. Secondly, from the
definition~\eqref{eq:xi_clamped} of $\xi^*$ and fig.~\ref{fig:Kraw} we deduce that there are
essentially three different behaviors:
\begin{enumerate}
\item $0\le\vep_F\le q$ (or equivalently $0\le\nu\le q$, since with our normalization $\nu=\vep$ in
  the thermodynamic limit)
  
  In this case $\rho(x,\vep_F)=0$ for $x\in[0,x_1(\vep_F)]\cup[x_2(\vep_F),\ell]$, i.e., there are two
  depletion intervals at each end of the chain.
  
\item $q\le\vep_F\le 1-q$

  Now $\rho(x,\vep_F)=1$ for $x\in[0,x_1(\vep_F)]$ and $\rho(x,\vep_F)=0$ for $x\in[x_2(\vep_F),\ell]$. In
  other words, there is a \emph{saturation interval} at the left end of the chain and a depletion
  interval at the right end.

\item $1-q\le\vep_F\le 1$

  In this case $\rho(x,\vep_F)=1$ for $x\in[0,x_1(\vep_F)]\cup[x_2(\vep_F),\ell]$. In other words,
  there are two saturation intervals at both ends of the chain.
    
\end{enumerate}

\begin{remark}
  The case $q=1/2$ is somewhat special, since then $q=1-q$ and the magnetic field is constant:
  $B(x)=q=1/2$. Now the range $q\le\vep_F\le 1-q$ reduces to the single value $\vep_F=\nu=1/2$
  (half filling), for which $\xi(x,1/2)=0$. Hence in this case
  \[
    a\rho(x,1/2)=\frac12,\qquad\all x\in[0,\ell],
  \]
  as expected.\qed
\end{remark}

  The Krawtchouk chain possesses several symmetries that we shall now briefly review. First of
  all, the chain parameters $J_n$ and $B_n$ behave under reflection about the chain's midpoint as
  \[
    J_{n}\mapsto J_{N-n-2}=J_n,\qquad B_{n}\mapsto B_{N-n-1}=N-1-B_n.
  \]
  Using these relations, it is straightforward to show that\footnote{Recall that in this case
    (with the normalization~\eqref{eq:JBKraw}) the chain energies are the numbers $\vep_k=k$, with
    $k=0,\dots,N-1$}
  \begin{equation}\label{eq:ref:Kraw}
    \phi_n(k)=(-1)^n\phi_{N-n-1}(N-1-k).
  \end{equation}
  This symmetry of the wave functions translates into a corresponding symmetry of the fermionic
  density. More precisely, for a Fermi energy $\vep_F=M-1$ (or filling fraction $M/N$), from the
  last equation we have
  \begin{align*}
    1&=\sum_{k=0}^{N-1}\phi_n^2(k)=\sum_{k=0}^{M-1}\phi_n^2(k)+\sum_{k=M}^{N-1}\phi_n^2(k)
       =\sum_{k=0}^{M-1}\phi_n^2(k)+\sum_{k=M}^{N-1}\phi_{N-n-1}^2(N-k-1)\\
     &=\sum_{k=0}^{M-1}\phi_n^2(k)+\sum_{k=0}^{N-M-1}\phi_{N-n-1}^2(k)
       =\mel{M}{c_n^\dagger c_n}{M}+\mel{N-M}{c_{N-n-1}^\dagger c_{N-n-1}}{N-M}.
  \end{align*}
  In terms of the continuum fermionic density $\rho(x,\vep_F)$, the previous relation
  becomes
  \begin{equation}
    \label{eq:ref:Kra}
    a\rho(x,\vep_F)=1-a\rho(\ell-x,1-\vep_F),
  \end{equation}
  where we have taken into account that in the continuum limit the energies are rescaled by a
  factor $N^{-1}$ (cf.~\eqref{eq:vepKraw}). To check that this symmetry is shared by the WKB
  approximation to the fermionic density, note first of all that $J(x)=J(\ell-x)$ and
  $B(\ell-x)=1-B(x)$, and therefore
  \[
    \xi(\ell-x,1-\vep_F)=-\xi(x,\vep_F).
  \]
  Since $x_{1,2}(1-\vep_F)=\ell-x_{2,1}(\vep_F)$, from eq.~\eqref{eq:rho:Kraw} we obtain
  \[
    a\rho(\ell-x,1-\vep_F)=\frac1\pi\arccos\xi(x,\vep_F)
    =\frac1\pi\big(\pi-\arccos\br(-\xi(x,\vep_F))\big)=1-a\rho(x,\vep_F),
  \]
  as claimed. In fact, the symmetry~\eqref{eq:ref:Kra} is apparent from the plots presented in
  fig.~\ref{fig:Kraw:rho} (left column).

  Likewise, under the parameter reflection $q\mapsto 1-q$ we have
  \[
    J_n\mapsto J_n,\qquad B_n\mapsto N-1-B_n,
  \]
  From these equations it is straightforward to derive the following relation between the
  single-particle wave functions with parameters $q$ and $1-q$:
  \[
    \phi_n(k;1-q)=(-1)^n\phi_n(N-1-k;q),
  \]
  which combined with eq.~\eqref{eq:ref:Kraw} yields the identity
  \[
    \phi_n(k;1-q)=\phi_{N-n-1}(k;q).
  \]
  In the continuum limit (after rescaling of the energies by $N^{-1}$), this relation yields
  \begin{equation}\label{eq:q1mq}
    \rho(x,\vep_F;1-q)=\rho(\ell-x,\vep_F;q).
  \end{equation}
  To show that the last equation is satisfied by the WKB approximation to the fermionic density,
  it suffices to note that
  \[
    J(x;1-q)=J(\ell-x;q),\qquad B(x;1-q)=B(\ell-x,q),
  \]
  and therefore $\xi(x,\vep;1-q)=\xi(\ell-x,\vep;q)$. Equation~\eqref{eq:q1mq} then follows
  immediately from eq.~\eqref{eq:rho:Kraw}, taking into account that
  $x_{1,2}(\vep_F;1-q)=\ell-x_{2,1}(\vep_F;q)$.

  \subsection{The rainbow chain}\label{subsec:rainbow}

  The rainbow chain was introduced in 2010 by Vitagliano and collaborators~\cite{VRL10} to
  illustrate violations of the area law, and has since been extensively
  studied~\cite{RRS14,RRS15,RDRCS17,TRS18,MLNR19,Sz25}. We shall take the model's parameters as
  \[
    J_n=\frac12\exp\br(-h\br|\frac12-\frac{n}N+\frac1{2N}\,\de_{n,N/2-1}|\,),\qquad
    B_n=0,
  \]
  where $h>0$ and $N$ is even. The continuum limit of these parameters is therefore
  \[
    J(x)=\frac12\,\e^{-h\br|\frac12-\frac{x}\ell|},\qquad B(x)=0,
  \]
  so that, by eq.~\eqref{eq:spec:bounds}, in the thermodynamic limit the single-particle energies
  should lie in the interval $[-1,1]$.

  We shall start by evaluating the approximation to the single-particle level density $D(\vep)$ in
  eq.~\eqref{eq:spacing}. By the particle-hole symmetry discussed in remark~\ref{rem:part-hole},
  we can restrict ourselves without loss of generality to negative energies. From the graph of
  $\pm 2J(x)$ (see fig.~\ref{fig:2J:rainb}) it is then apparent that there are essentially two
  regimes:
  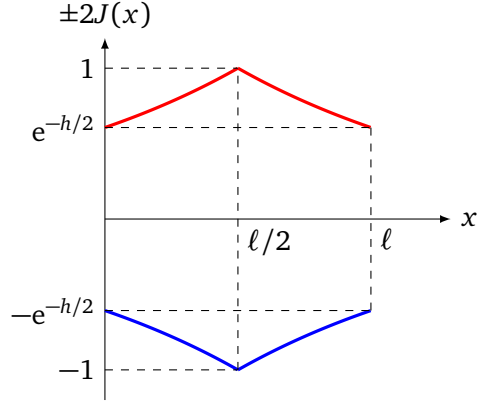
\begin{figure}[t!]
    \centering
    \begin{tikzpicture}[xscale=3.5,yscale=2]
    \draw[-latex] (0,0) -- (1.3,0) node[right] {$x$};
    \draw[-latex] (0,-1.2) -- (0,1.2) node[above] {$\pm2J(x)$};
    \draw[very thick,domain=0:.5,red,samples=300] plot (\x,{exp(-(1/2-\x))});
    \draw[very thick,domain=.5:1,red,samples=300] plot (\x,{exp((1/2-\x))});
    \draw[very thick,domain=0:.5,blue,samples=300] plot (\x,{-exp(-(1/2-\x))});
    \draw[very thick,domain=.5:1,blue,samples=300] plot (\x,{-exp((1/2-\x))});
    \draw[dashed] (0,{exp(-1/2)}) node[left] {$\e^{-h/2}$} -- (1,{exp(-1/2)});
    \draw[dashed] (0,{-exp(-1/2)}) node[left] {$-\e^{-h/2}$} -- (1,{-exp(-1/2)});
    \draw[dashed] (1,{-exp(-1/2)}) -- (1,{exp(-1/2)});
    \node[anchor=north west] at (1,0) {$\ell$};
    \draw[dashed] (1/2,-1) -- (1/2,1);
    \node[anchor=north west] at (1/2,0) {$\ell/2$};
    \draw[dashed] (0,1) node[left] {$1$} -- (1/2,1);
    \draw[dashed] (0,-1) node[left] {$-1$} -- (1/2,-1);
    \end{tikzpicture}
    \caption{Schematic plot of $\pm 2J(x)$ for the rainbow chain.}
    \label{fig:2J:rainb}
  \end{figure}
  \begin{enumerate}
  \item $-1\le\vep\le-\e^{-h/2}$

    In this case $\xi(x,\vep)<-1$ for $x\in(0,x_1(\vep))\cup(x_2(\vep),\ell)$, where the
    turning points $x_{1,2}(\vep)$ are given by
    \begin{equation}
      \label{eq:x12:rainb}
      x_{1,2}(\vep)=\frac{\ell}2\pm\frac{\ell}h\,\log\abs{\vep}.
    \end{equation}
    From eq.~\eqref{eq:A2vep} and the symmetry of $J(x)$ under reflections about $x=\ell/2$ it
    then follows that
    \begin{equation*}
      A(\vep)^{-2}=2\int_{x_1(\vep)}^{\ell/2}\frac{\dd x}{\sqrt{4J^2(x)-\vep^2}}
      =2\ell\int_{x_1(\vep)/\ell}^{1/2}\frac{\dd x}{\sqrt{\e^{-h(1-2x)}-\vep^2}}
      =\frac\ell{h}\int_0^{-2\log\abs{\vep}}\frac{\dd s}{\sqrt{\e^{-s}-\vep^2}}.
    \end{equation*}
    Evaluating the integral we thus obtain
    \begin{equation}
      \label{eq:Dep:rainb}
      D(\vep)=\frac1{\pi a}A(\vep)^{-2}
      =\frac{\ell}{ha\abs{\vep}}\,\br(1-\frac{2}{\pi}\arcsin\abs{\vep}),\qquad
      -1\le\vep\le-\e^{-h/2}.
    \end{equation}

  \item  $-\e^{-h/2}\le\vep\le0$

    Now $\abs{\xi(x,\vep)}\le1$ for all $x\in[0,\ell]$, and therefore
    \begin{align*}
      A(\vep)^{-2}&=2\int_{0}^{\ell/2}\frac{\dd x}{\sqrt{4J^2(x)-\vep^2}}
                    =\frac{\ell}{h}\int_0^h\frac{\dd s}{\sqrt{\e^{-s}-\vep^2}},
    \end{align*}
    and hence
    \begin{equation}
      \label{eq:Dep:rainb2}
      D(\vep)=\frac1{\pi a}A(\vep)^{-2}
      =\frac{2\ell}{\pi ha\abs{\vep}}\,\br(\arcsin\br(\e^{h/2}\abs{\vep})
                    -\arcsin\abs{\vep}),\qquad
      -\e^{-h/2}\le\vep\le0.
    \end{equation}
  \end{enumerate}
  Combining the value of $A(\vep)$ just computed with eq.~\eqref{eq:envelope} we obtain the
  following explicit equation for the envelopes of the single-particle approximate eigenfunctions:
  \begin{equation}\label{eq:env:rainb}
    y=\pm\sqrt{\frac{h\abs{\vep}}{\ell}}\,\br(\e^{-h\br\abs{1-\frac{2x}\ell}}-\vep^2)^{-1/4}
    \begin{cases}
      \br(\frac\pi2-\arcsin\abs{\vep})^{-1/2},
      &\e^{-h/2}\le\abs{\vep}\le1\\
      \Big(\arcsin\br(\e^{h/2}\abs{\vep})-\arcsin\abs{\vep}\Big)^{-1/2},
      & \abs{\vep}\le\e^{-h/2},                                                       
    \end{cases}
  \end{equation}
    \begin{figure}[t!]
    \includegraphics[width=.48\textwidth]{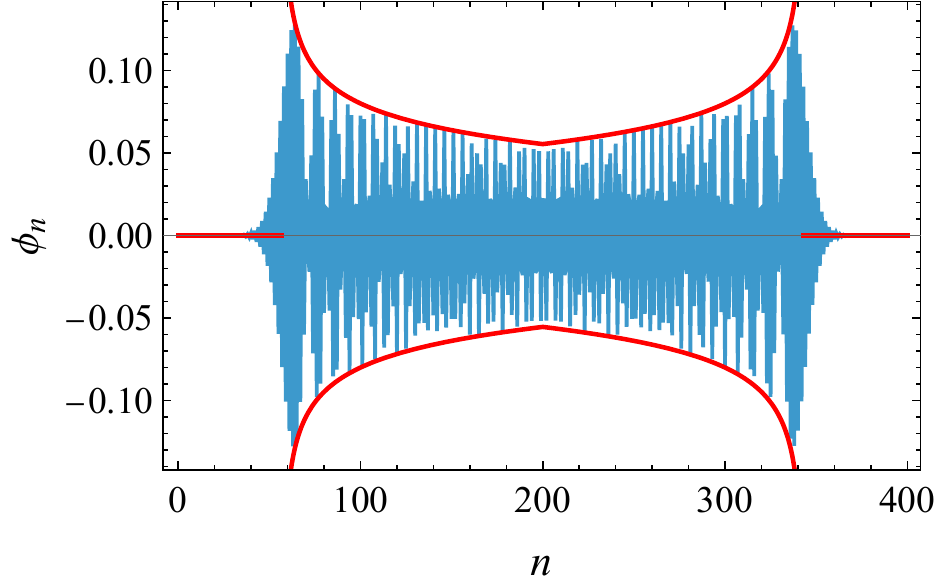}\hfill
    \includegraphics[width=.48\textwidth]{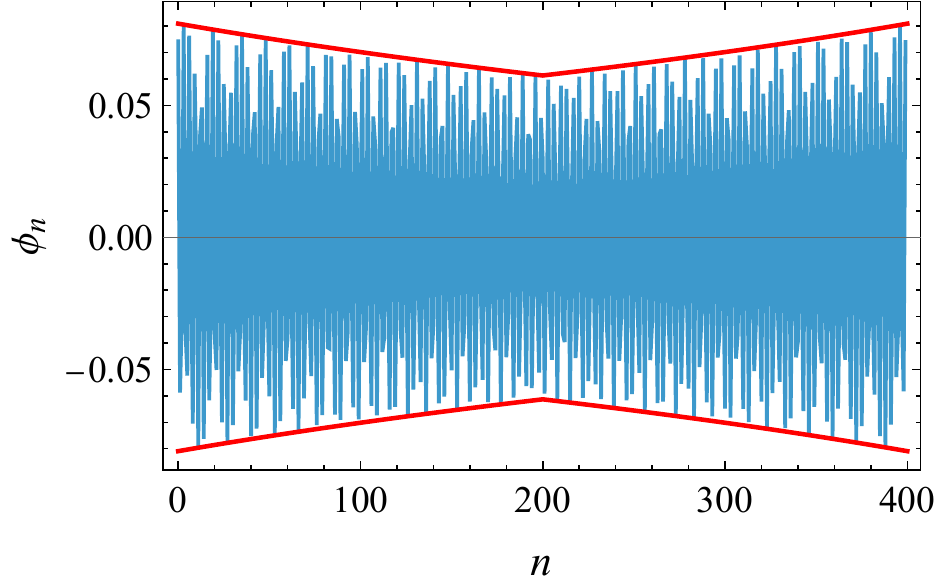}
    \caption{Single-particle eigenfunctions $\phi_n(\vep)$ with energies $\vep=N/8$ (left) and
      $\vep=2N/5$ (right) for the rainbow chain with $h=1$, $N=400$, compared to the
      envelopes~\eqref{eq:env:rainb} of the corresponding approximate WKB wave functions (red
      lines).}
    \label{fig:phi:rainb}
  \end{figure}%
  where in the first line it is understood that $y=0$ for $x\notin[x_1(\vep),x_2(\vep)]$. As an
  example, in fig.~\ref{fig:phi:rainb} we show these envelopes together with the
  graphs of the corresponding numerical single-particle eigenfunctions for a chain of $N=400$
  spins with parameter $h=1$, at the energies $\vep_{N/8}$ and $\vep_{2N/5}$. In particular, it is
  apparent from these plots that $y(x)=y(\ell-x)$, which is a direct consequence of the symmetry
  of the coupling $J(x)$ about the chain's midpoint. (This is also obvious from
  eq.~\eqref{eq:env:rainb}.)
  
  We shall next evaluate the approximation~\eqref{eq:filling} to the filling fraction
  $\nu(\vep_F)$ at a Fermi energy $\vep_F$, supposing again without loss of generality that
  $\vep_F\le0$. If $-1\le\vep_F\le-\e^{-h/2}$ then $\xi^*(x,\vep_F)=-1$ outside the interval
  $(x_1(\vep_F),x_2(\vep_F))$, and therefore (again by symmetry about $x=\ell/2$)
  \begin{align*}
    \nu(\vep_F)
    &=\frac{2}{\pi\ell}\int_{x_1(\vep_F)}^{\ell/2}\dd x\,\arccos(-\xi(x,\vep_F))
      =\frac{2}{\pi}\int_{x_1(\vep_F)/\ell}^{1/2}\dd x\arccos\br(\e^{h\br(\frac12-x)}\abs{\vep_F})\\[2mm]
    &=\frac2{\pi h}\,\int_{\abs{\vep_F}}^1\frac{\dd s}{s}\,\arccos s
      =\frac2{\pi h}\,\int_{\abs{\vep_F}}^1\frac{\dd s}{s}\,\br(\frac\pi2-\arcsin s)
      =-\frac1h\log\abs{\vep_F}-\frac2{\pi h}\,f(\abs{\vep_F}),
  \end{align*}
  where
  \[
    f(t)\coloneqq\int_t^1\frac{\dd s}s\,\arcsin s.
  \]
  In fact, the RHS can be evaluated in closed form in terms of Clausen's integral
  \[
    \Cl_2(\th)\coloneqq-\int_0^\th\dd x\,\log\br(2\sin\tfrac
    x2)=\sum_{n=1}^\infty\frac{\sin(n\th)}{n^2},\qquad 0<\th<2\pi,
  \]
  which is in turn related to the standard dilogarithm function~\cite{OLBC10} by
  \[
    \Cl_2(\th)=\Im\Li_2\!\br(\e^{\iu\th})=
    \frac1{2i}\br(\Li_2\!\br(\e^{\iu\th})-\Li_2\!\br(\e^{-\iu\th})).
  \]
  More precisely,
  \[
    f(t)=\frac{\pi}2\log 2-\log(2t)\arcsin t-\frac12\Cl_2(2\arcsin t),
  \]
  and hence
  \begin{multline}
    \label{eq:nu:rainb}
    \nu(\vep_F)=-\frac1h\log\abs{2\vep_F}\\+\frac{1}{\pi
      h}\br\Big(2\log\abs{2\vep_F}\arcsin\abs{\vep_F}
    +\Cl_2(2\arcsin\abs{\vep_F})),\quad
    -1\le\vep_F\le-\e^{-h/2}.
  \end{multline}

\begin{figure}[t!]
    \includegraphics[width=7cm]{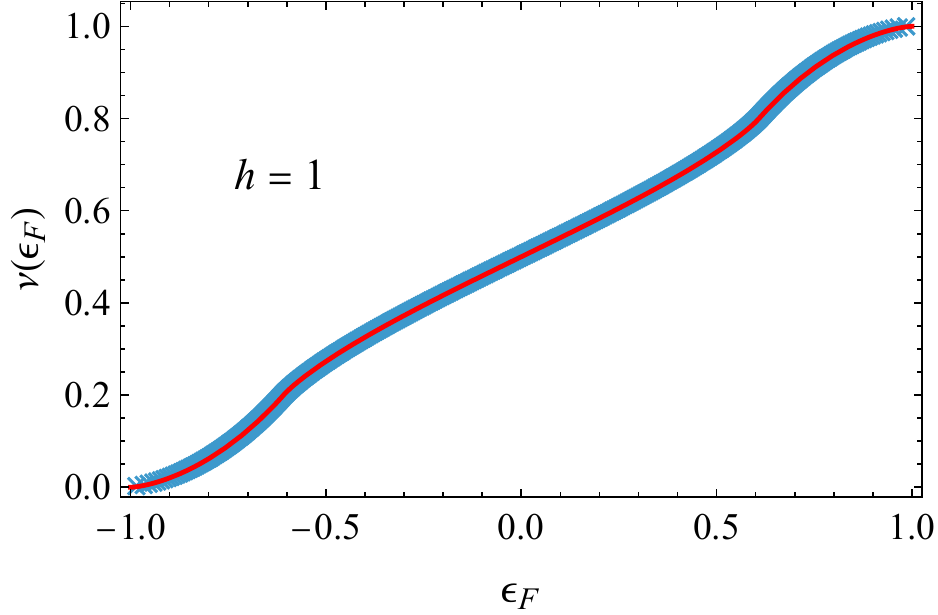}\hfill
    \includegraphics[width=7cm]{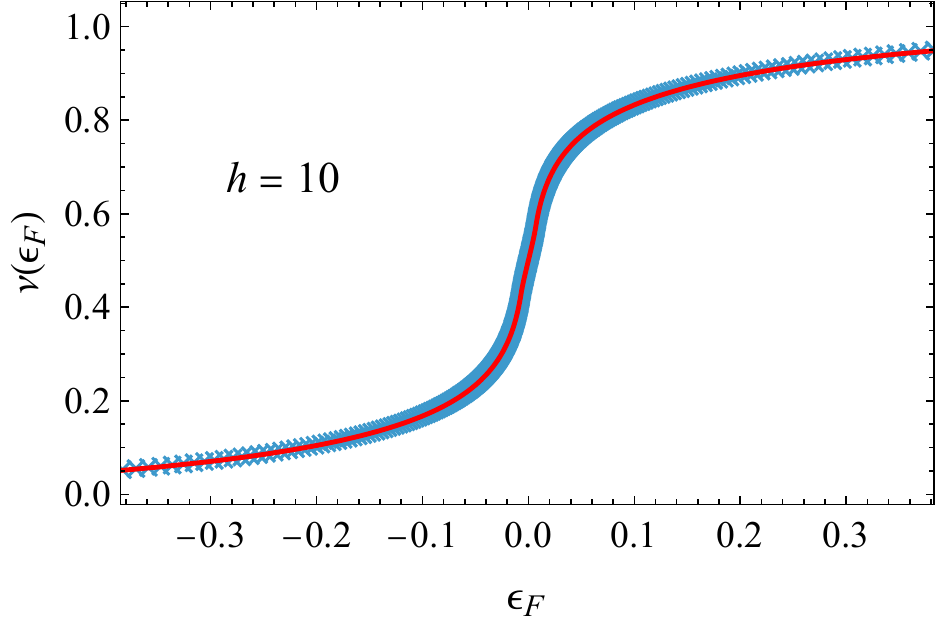}\hfill
    \caption{Filling fraction $\nu$ vs.~Fermi energy~$\vep_F$ for the rainbow chain with $N=400$
      spins and parameters $h=1,10$ (blue markers), compared to their WKB
      approximation~\eqref{eq:nu:rainb}-\eqref{eq:nu:rainb2} (red line).}
    \label{fig:nu:rainb}
  \end{figure}

  Consider next the case $-\e^{-h/2}\le\vep_F\le0$, in which $\abs{\xi(x,\vep_F)}\le 1$ for all
  $x\in[0,\ell]$. Now
  \begin{align}
    \nu(\vep_F)
    &=\frac{2}{\pi\ell}\int_{0}^{\ell/2}\dd x\,\arccos(-\xi(x,\vep_F))
      =\frac{2}{\pi}\int_{0}^{1/2}\dd x\arccos\br(\e^{h\br(\frac12-x)}\abs{\vep_F})\notag\\
    &=\frac2{\pi h}\,\int_{\abs{\vep_F}}^{\e^{h/2}\abs{\vep_F}}\frac{\dd s}{s}\,\arccos s
      =\frac2{\pi h}\,\int_{\abs{\vep_F}}^{\e^{h/2}\abs{\vep_F}}\frac{\dd s}{s}\,
      \br(\frac\pi2-\arcsin s)\notag\\
    &=\frac12+\frac2{\pi h}\br(f\br(\e^{h/2}\abs{\vep_F})-f(\abs{\vep_F})).
      \label{eq:nu:rainb2}
  \end{align}
  In fact, our numerical calculations show that the
  approximation~\eqref{eq:nu:rainb}-\eqref{eq:nu:rainb2} (extended to positive energies using
  eq.~\eqref{eq:symm:rhonu}) is remarkably accurate across the full energy range
  $\vep_F\in[-1,1]$; see, e.g., fig.~\ref{fig:nu:rainb} for the cases $h=1$ and $h=10$.

  We shall finally discuss the approximation to the fermionic density given by
  eq.~\eqref{eq:occupation_density}. From fig.~\ref{fig:2J:rainb} it is clear that in the range
  $-1\le\vep_F\le -\e^{-h/2}$ we have $\arccos(-\xi^*(\vep_F))=0$ for
  $x\notin(x_1(\vep_F),x_2(\vep_F))$. Hence in this case there are two depletion intervals at the
  two ends of the chain, limited by the turning points $x_{1,2}(\vep_F)$, while in the interval
  $[x_1(\vep_F),x_2(\vep_F)]$ we have
  \begin{equation}
    \label{eq:rho:rainb}
    a\rho(x,\vep_F)=\frac1{\pi}\,\arccos\!\br(\abs{\vep_F}\e^{h\br|\frac12-\frac{x}\ell|}).
  \end{equation}
  Note that $\rho(x,\vep_F)$ is clearly symmetric about $x=\ell/2$, as expected by the analogous
  symmetry of $J(x)$. On the other hand, for $-\e^{-h/2}\le\vep_F\le0$ there are no depletion
  intervals, since $\abs{\xi(x,\vep_F)}\le1$ for all $x\in[0,\ell]$, so that the last equation is
  valid in the whole interval $[0,\ell]$. Of course, the fermionic density for positive energies
  can be computed from the symmetry relation~\eqref{eq:symm:rhonu}.

   \begin{figure}[t!]
    \includegraphics[width=.48\textwidth]{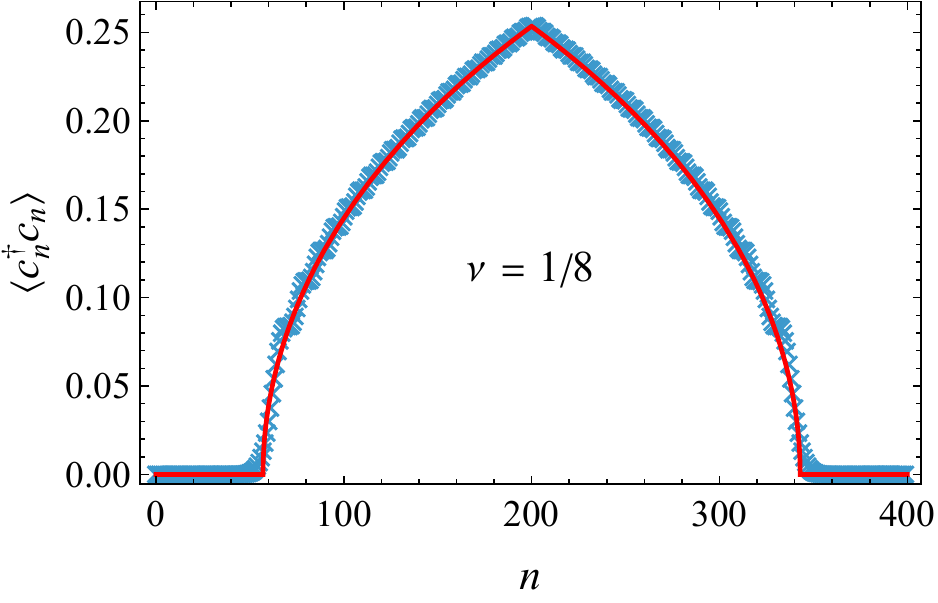}\hfill
    \includegraphics[width=.48\textwidth]{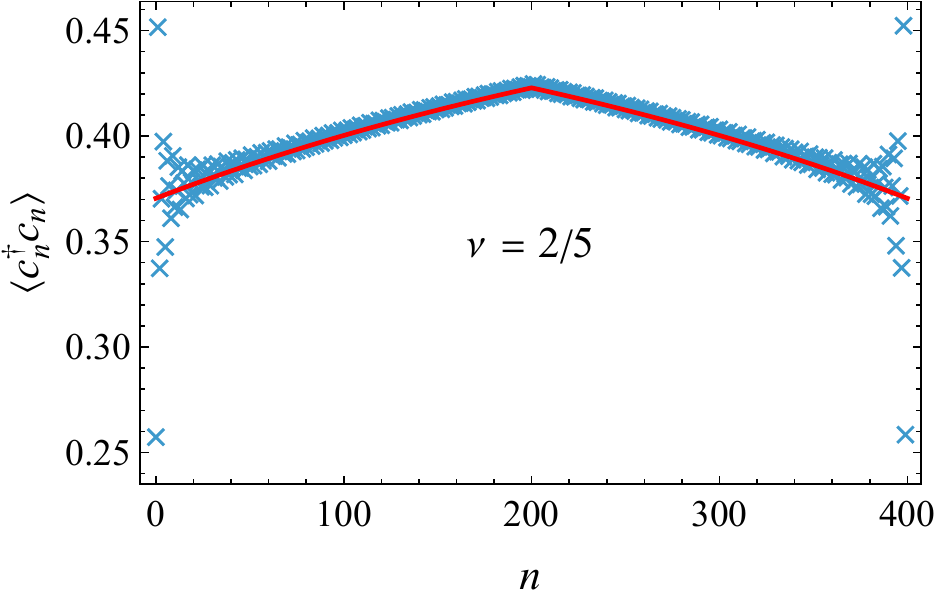}
    \caption{Local fermion density for the rainbow chain with $h=1$, $N=400$, and filling
      fractions $\nu=1/8$ and $2/5$ (blue markers) compared to its asymptotic
      approximation~\eqref{eq:rho:rainb} (red line).}
    \label{fig:rho:rainb}
  \end{figure}

  In fig.~\ref{fig:rho:rainb} we have plotted the fermionic density of the rainbow chain with
  $N=400$ spins and parameter $h=1$ for the fillings $\nu=1/8$ and $\nu=2/5$, compared to its
  approximation~\eqref{eq:rho:rainb}. For these values of $N$ and $h$ we have
  $\e^{-h/2}=0.60653\dots$, so that the corresponding Fermi energies satisfy
  $\vep_F(1/8)=-0.69945<-\e^{-h/2}$ and $\vep_F(2/5)=-0.24004>-\e^{-h/2}$, respectively. Thus for
  $\nu=1/8$ there are two depletion intervals, namely (by eq.~\eqref{eq:x12:rainb}) $[0,57.018]$
  and $[342.982,400]$, while for $\nu=2/5$ there are none. As can be seen from
  fig.~\ref{fig:rho:rainb}, this prediction is borne out by the numerical data. In fact, the
  approximation~\eqref{eq:rho:rainb} is reasonably accurate for both values of $\nu$, although for
  $\nu=2/5$ characteristic Friedel oscillations appear at the chain's ends.

  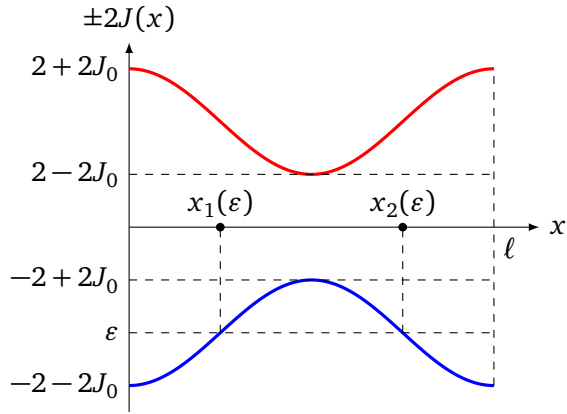
\begin{figure}[b!]
  \centering
    \begin{tikzpicture}[xscale=1.2,yscale=.7]
    \draw[-latex] (0,0) -- (4.5,0) node[right] {$x$};
    \draw[-latex] (0,-3.5) -- (0,3.5) node[above] {$\pm2J(x)$};
    \draw[very thick,domain=0:4,samples=400,red] plot
    (\x,{2*(1+.5*cos(90*\x))});
    \draw[very thick,domain=0:4,samples=400,blue] plot
    (\x,{-2*(1+.5*cos(90*\x))});
    \draw[dashed] (0,-2) node[left] {$\vep$} -- (4,-2);
    \filldraw[cm={1/1.2,0,0,1/.7,(1,0)}] (0,0) circle (0.05);
    \filldraw[cm={1/1.2,0,0,1/.7,(3,0)}] (0,0) circle (0.05);
    \draw[dashed] (1,-2) -- (1,0) node [above] {$x_1(\vep)$};
    \draw[dashed] (3,-2) -- (3,0) node [above] {$x_2(\vep)$};
    \draw[dashed] (4,-3) -- (4,0) node [anchor=north west] {$\ell$};
    \draw[dashed] (4,0) -- (4,3);
    \node[left] at (0,3) {$2+2J_0$};
    \node[left] at (0,-3) {$-2-2J_0$};
    \draw[dashed] (0,1) node[left] {$2-2J_0$}-- (4,1);
    \draw[dashed] (0,-1) node[left] {$-2+2J_0$}-- (4,-1);
  \end{tikzpicture}
  \caption{Plot of $\pm 2J(x)$ for the cosine chain~\eqref{eq:cos_chain}, showing the turning
    points $x_{1,2}(\vep)$ for the specified energy $\vep$.}
  \label{fig:cos}
\end{figure}%

\subsection{The cosine chain}

\begin{figure}[t!]
  \centering
    \includegraphics[height=.33\textwidth]{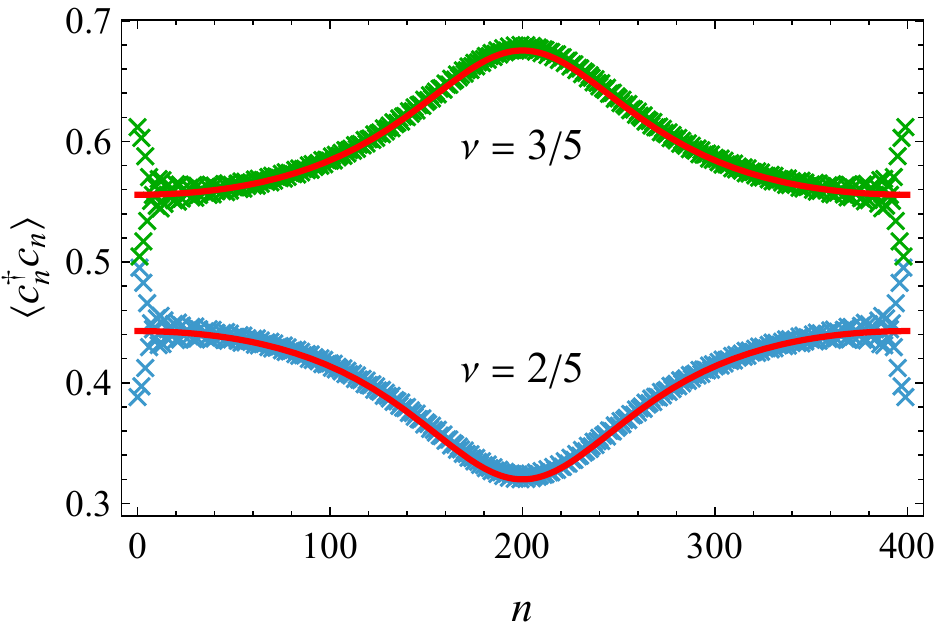}\hfill
  \includegraphics[height=.33\textwidth]{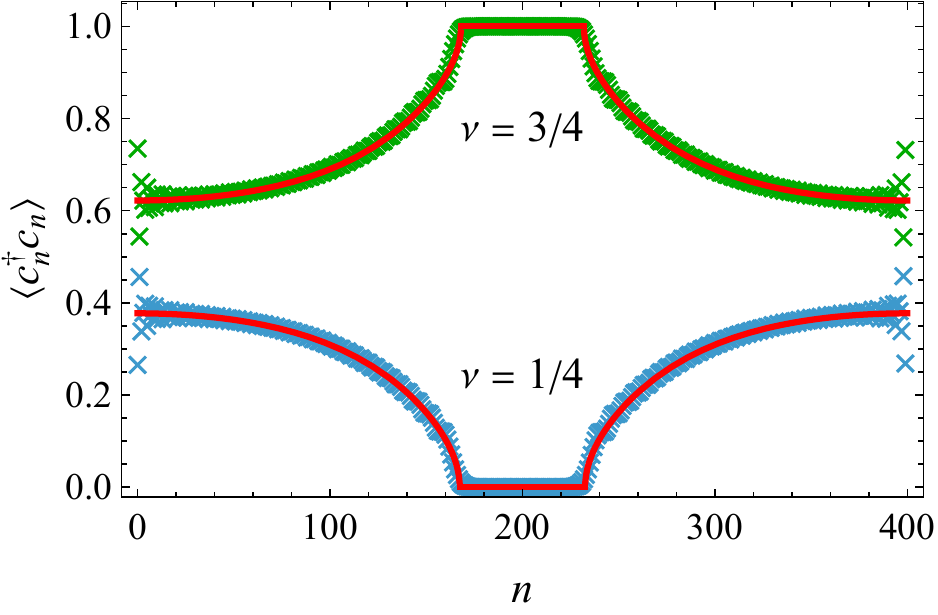}
  \caption{Local fermion density for the cosine chain~\eqref{eq:cos_chain:J} with $J_0=1/2$,
    $N=400$ spins, and several values of the filling $\nu$ (blue and green markers) compared to
    their WKB approximation~\eqref{eq:occupation_density} (continuous red line).}
  \label{fig:cos_chain}
\end{figure}

Consider next the cosine chain introduced in ref.~\cite{MSSSR22}, with parameters
\begin{equation}\label{eq:cos_chain:J}
  J_n=1+J_0\cos\br(\frac{2\pi n}N),\qquad B_n=0,
\end{equation}
whose continuum limit is given by
\begin{equation}
  \label{eq:cos_chain}
  J(x)=1+J_0\cos\br(\frac{2\pi x}\ell),\qquad B(x)=0.
\end{equation}
Note that this chain's hoppings are \emph{not} symmetric about its midpoint (i.e., $J_n$ is not
invariant under the reflection $n\mapsto N-2-n$), although their continuum
counterparts~\eqref{eq:cos_chain} are. By eq.~\eqref{eq:spec:bounds}, the single-particle spectrum
lies inside the interval $[-2-2J_0,2+2J_0]$. It is apparent from fig.~\ref{fig:cos} that the
qualitative behavior of the fermionic density depends on whether the Fermi energy $\vep_F$ is
smaller or larger in absolute value than $2-2J_0$. Indeed, in the former case, i.e., for
$-2+2J_0\le\vep_F\le 2-2J_0$, there are no depletion/saturation regions, and the fermionic density
is given by
\begin{equation}\label{eq:rho:cos}
  \rho(x,\vep_F)=\frac1{\pi a}\,\arccos\br(\frac{\abs{\vep_F}/2}{1+J_0\cos\br(\frac{2\pi
      x}\ell)})
\end{equation}
for all $x\in[0,\ell]$. For instance, for $\vep_F\in[-2+2J_0,0]$, $\rho$ decreases monotonically
over the interval $[0,\ell/2]$ from its maximum value
$\rho(0,\vep_F)=(\pi a)^{-1}\arccos\br(\frac{\abs{\vep_F}}{2(1+J_0)})$ to its minimum
$\rho(\ell/2,\vep_F)=(\pi a)^{-1}\arccos\br(\frac{\abs{\vep_F}}{2(1-J_0)})$, and is symmetric
about $\ell/2$ (see, e.g., fig.~\ref{fig:cos_chain}, left panel, with $\nu=2/5$, for which
$\vep_F=-0.53597\in[-2+2J_0,0]=[-1,0]$). The behavior of $\rho$ for $\vep_F\in[0,2-2J_0]$ is
completely analogous (see fig.~\ref{fig:cos_chain}, left panel, with $\nu=3/5$, for which
$\vep_F=0.52343\in[0,2-2J_0]=[0,1]$). A special case occurs when $\vep_F=0$ (i.e., at half
filling), when $\rho(x,0)=1/(2a)$ becomes constant. On the other hand, for $\vep_F$ in the range
$[-2-2J_0,-2+2J_0]$ or $[2-2J_0,2+2J_0]$ there are two turning points $x_{1,2}(\vep_F)$ given by
\begin{equation}\label{eq:cos_TP}
  x_1(\vep_F)=\frac\ell{2\pi}\arccos\br(\frac{\abs{\vep_F}-2}{2J_0}),\qquad
  x_2(\vep_F)=\ell-x_1(\vep_F)
\end{equation}
(cf.~fig.~\ref{fig:cos}).
\begin{figure}[b!]
  \centering
  \includegraphics[height=.31\textwidth]{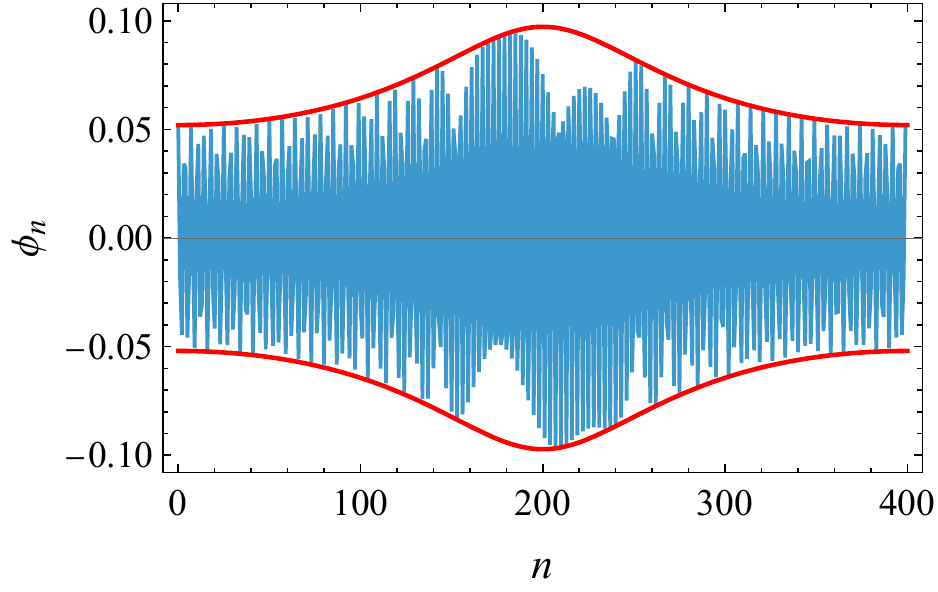}\hfill
  \includegraphics[height=.31\textwidth]{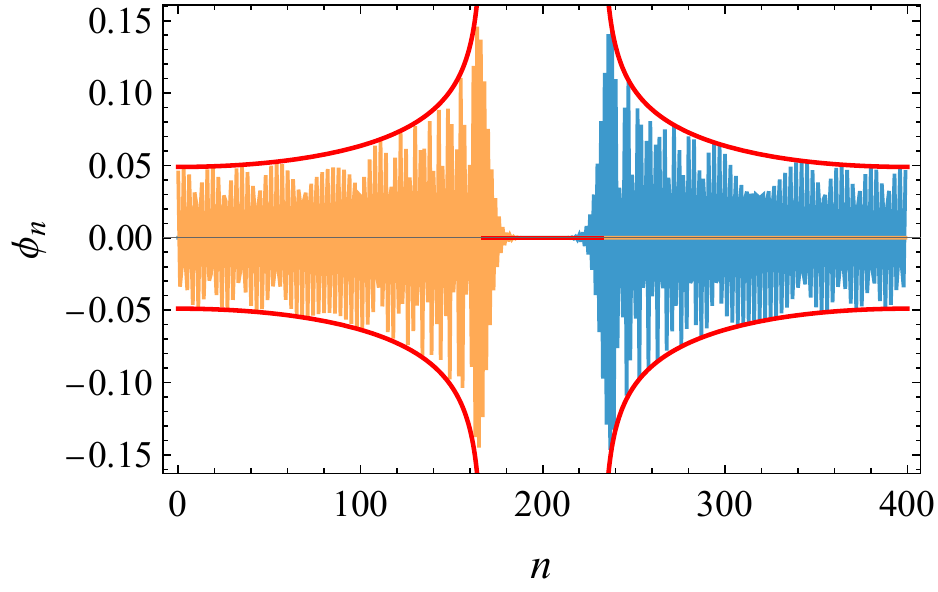}
  \caption{WKB envelopes~\eqref{eq:envelope} for the cosine chain~\eqref{eq:cos_chain:J} with
    $J_0=1/2$ and $N=400$ spins (red lines), compared to the exact wave function
    $\phi_n(\vep_{2N/5})$ (left) and the linear combination with equal coefficients $1/\sqrt2$ of
    the exact eigenfunctions $\phi_n(\vep_{N/4})$ (right panel, blue) and
    $\phi_n(\vep_{\frac N4+1})$ (right panel, orange).}
  \label{fig:phi-Cos}
\end{figure}%
In this case $[x_1(\vep_F),x_2(\vep_F)]$ is a depletion interval for $\vep_F\in [-2-2J_0,-2+2J_0]$
and a saturation interval for $\vep_F\in [2-2J_0,2+2J_0$]. Outside this interval, the fermionic
density is again given by eq.~\eqref{eq:rho:cos}; see, e.g., fig.~\ref{fig:cos_chain}, right
panel, with filling fractions corresponding to the energies $-1.12833$ and $1.12323$.

We have been able to obtain closed-form expressions for the normalization constant
$A(\vep)$---and, hence, for the equation~\eqref{eq:envelope} of the WKB envelopes---in terms of a
complete elliptic integral of the first kind, but we shall omit them here due to their excessive
complexity. More precisely, from the discussion in section~\ref{sec:WKB} and
remark~\ref{rem:eff_wf}, it follows that when there is no depletion/saturation, i.e., for
$\abs{\vep}<2-2J_0$, these curves are genuine envelopes for the corresponding eigenfunctions,
while for $\abs{\vep}\ge 2-2J_0$ they are the envelopes of the linear combinations
$(\phi_n(\vep_k)+\phi_n(\vep_{k+1}))/\sqrt2$ of two neighboring eigenfunctions
(cf.~fig.~\ref{fig:phi-Cos}).

\begin{figure}[t!]
  \centering
  \includegraphics[height=.33\textwidth]{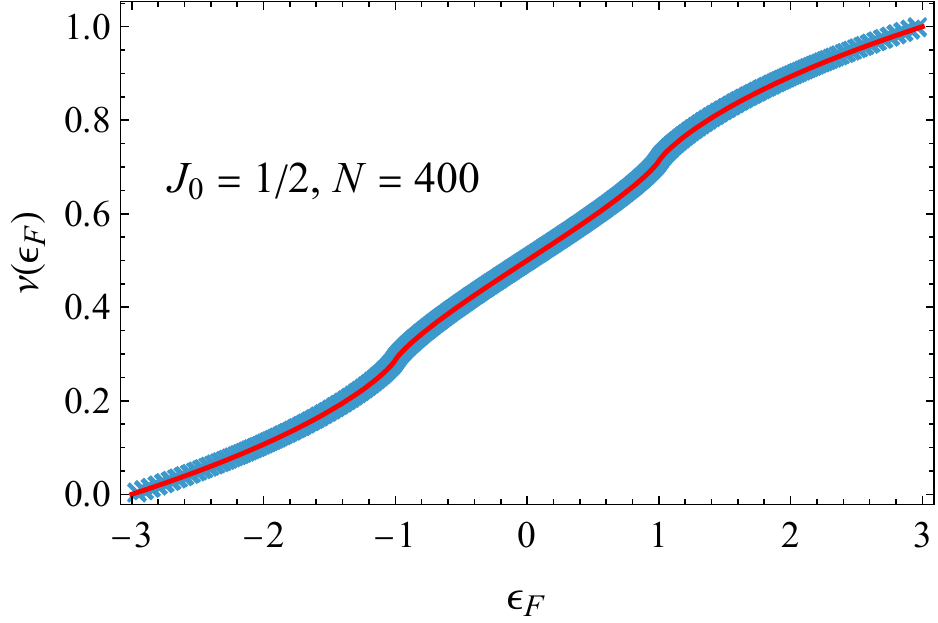}\hfill
  \includegraphics[height=.33\textwidth]{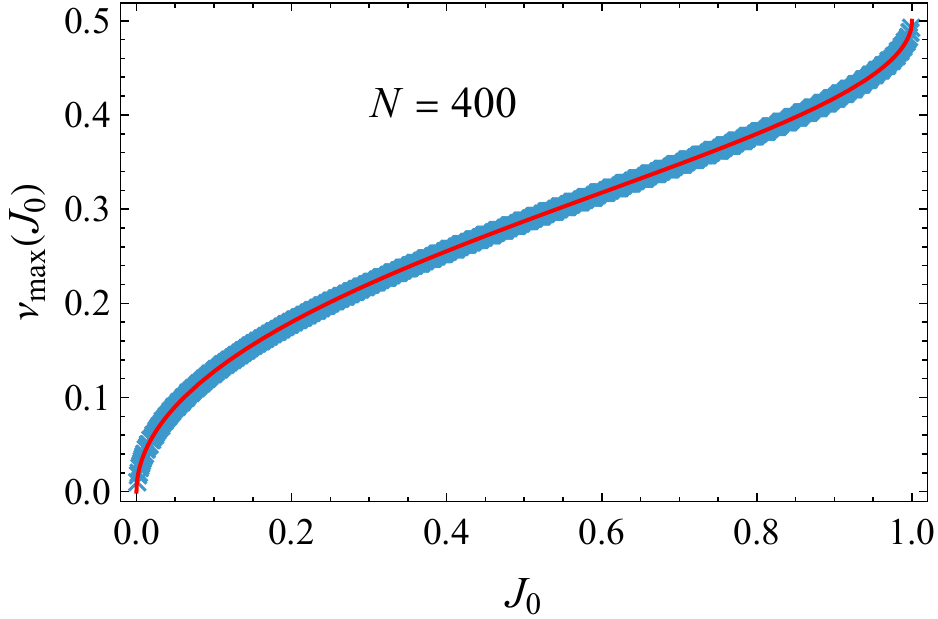}
  \caption{Left: filling fraction $\nu(\vep_F)$ for the cosine chain~\eqref{eq:cos_chain:J} with
    $J_0=1/2$ and $N=400$ spins (blue markers) compared to its WKB
    approximation~\eqref{eq:filling}. Right: maximum filling fraction $\nu_{\mathrm{max}}(J_0)$
    for which the fermionic density exhibits depletion for $N=400$ and
    $J_0=0.001,0.002,\dots,0.999$ (blue markers), compared to its WKB
    approximation~\eqref{eq:numax_Cos} (red line).}
  \label{fig:numax-Cos}
\end{figure}

Although we have not found an explicit expression for the
filling fraction $\nu(\vep_F)$ in terms of the Fermi energy $\vep_F$, it can of course be
evaluated numerically without difficulty for given $\vep_F$ and $J_0$ from eq.~\eqref{eq:filling}
(see, e.g., fig.~\ref{fig:numax-Cos} (left) for a plot of the filling fraction $\nu(\vep_F)$ for
$J_0=1/2$ and $N=400$ spins compared to its WKB approximation). In particular, the maximum filling
fraction $\nu_{\mathrm{max}}(J_0)$ for which the fermionic density features a depletion interval,
given by the integral
\begin{equation}\label{eq:numax_Cos}
  \nu_{\mathrm{max}}(J_0)=\frac1{\pi\ell}\int_{0}^\ell\dd x\,\arccos\br\big(-\xi(2J_0-2))=
  \frac1{\pi^2}\int_0^\pi\dd s\,\arccos\br(\frac{1 - J_0}{1 + J_0 \cos s}),
\end{equation}
can be computed in this way for any $J_0\in(0,1)$ (see fig.~\ref{fig:numax-Cos} (right) for the
graph of this function for $N=400$ spins).

\subsection{An asymmetric cosine chain}

As our last example, we shall consider the generalization of the cosine chain discussed above with
parameters
\begin{subequations}\label{eq:gen_cos_chain}
  \begin{equation}\label{eq:gen_cos_chain1}
    J_n=1+J_0\cos\br(\frac{2\pi r n}N),\qquad B_n=b\br(\frac{n}{N})^2,
  \end{equation}
  where for definiteness we have taken
  \begin{equation}\label{eq:cos_chain_2}
    J_0=\frac34\,,\qquad b=5,\qquad r=2.
  \end{equation}
\end{subequations}
The continuum version of the chain parameters is therefore given by
\begin{equation}\label{eq:gen_cos_chain_cont}
  J(x)=1+J_0\cos\br(\frac{2\pi r x}\ell),\qquad B(x)=b\br(\frac{x}{\ell})^2.
\end{equation}
Hence in this example neither the chain's parameters~\eqref{eq:gen_cos_chain} nor their continuum
counterparts \eqref{eq:gen_cos_chain_cont} are symmetric about the chain's midpoint. In
fig.~\ref{fig:gen_cos_chain} we have plotted the graphs of the functions $B(x)\pm2J(x)$ which, as
discussed above, determine the different qualitative behaviors of the fermionic density.
\begin{figure}[t!]
  \centering
  \begin{tikzpicture}[xscale=5.5,yscale=.8]
    \draw[-latex] (0,0) -- (1.1,0) node[right] {$x$};
    \draw[-latex] (0,-4) -- (0,9) node[above] {$B(x)\pm2J(x)$};
    \draw[very thick,red,domain=0:1,samples=400] plot
    (\x,{5*\x*\x+2*(1+.75*cos(720*\x))});
    \draw[very thick,blue,domain=0:1,samples=400] plot
    (\x,{5*\x*\x-2*(1+.75*cos(720*\x))});
    {\small
    \draw[dashed] (0,-3.5) node [left] {$e_0$} -- (1,-3.5);
    \draw[dashed] (0,-2.3009) node [left] {$e_1$} -- (1,-2.3009);
    \draw[dashed] (0,-0.173703) node [left] {$e_2$} -- (1,-0.173703);
    \draw[dashed] (0,0.799825) node [left] {$e_3$} -- (1,0.799825);
    \draw[dashed] (0,1.29292)  -- (1,1.29292);
    \node[left] at (0,1.24292){$e_4$};
    \draw[dashed] (0,1.5) -- (1,1.5);
    \node[left] at (0,1.55) {$e_5$};
    \draw[dashed] (0,2.43844) node [left] {$e_6$} -- (1,2.43844);
    \draw[dashed] (0,3.19719) node [left] {$e_7$} -- (1,3.19719);
    \draw[dashed] (0,3.5) node [left] {$e_8$} -- (1,3.5);
    \draw[dashed] (0,4.80547) node [left] {$e_9$} -- (1,4.80547);
    \draw[dashed] (0,8.5) node [left] {$e_{10}$} -- (1,8.5);
    }
    \draw[white] (0,-4) -- (0,-4.5);
    \draw[dashed] (1,-3.5) -- (1,0);
    \draw[dashed] (1,0) node [anchor=north west] {$\ell$}-- (1,8.5);
  \end{tikzpicture} \quad
  \includegraphics[height=10.5cm]{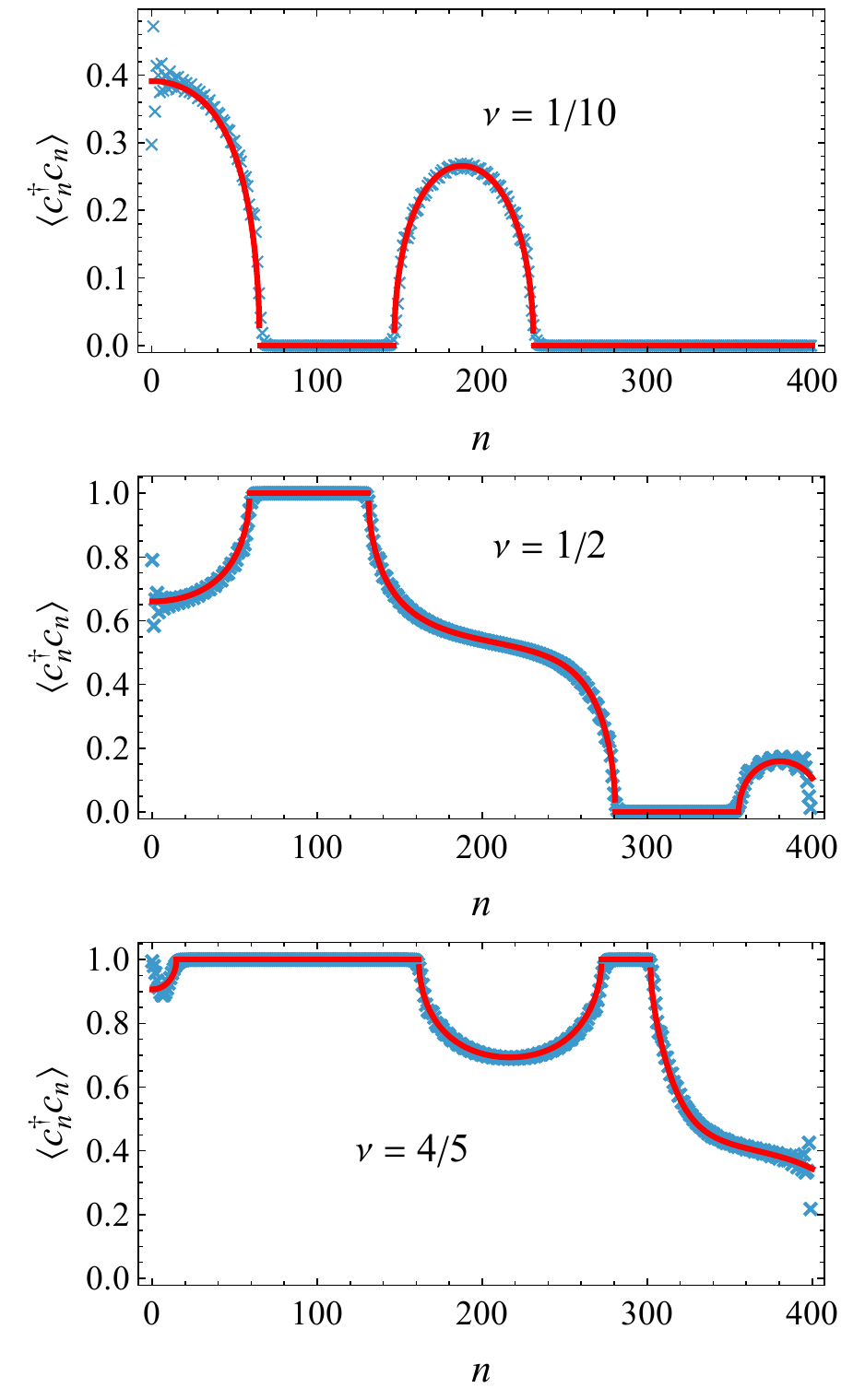}
  \caption{Left: plot of $B(x)\pm2J(x)$ for the general cosine chain~\eqref{eq:gen_cos_chain}. The
    qualitative behavior of the fermionic density is determined by the interval
    $[e_i,e_{i+1}]$ (with $i=0,1,\dots,9$) containing the Fermi energy $\vep_F$. Right: local
    fermion density for the chain~\eqref{eq:gen_cos_chain} with $N=400$ spins for several filling
    fractions (blue markers), compared to its WKB approximation~\eqref{eq:occupation_density} (red
    line).}
  \label{fig:gen_cos_chain}
\end{figure}
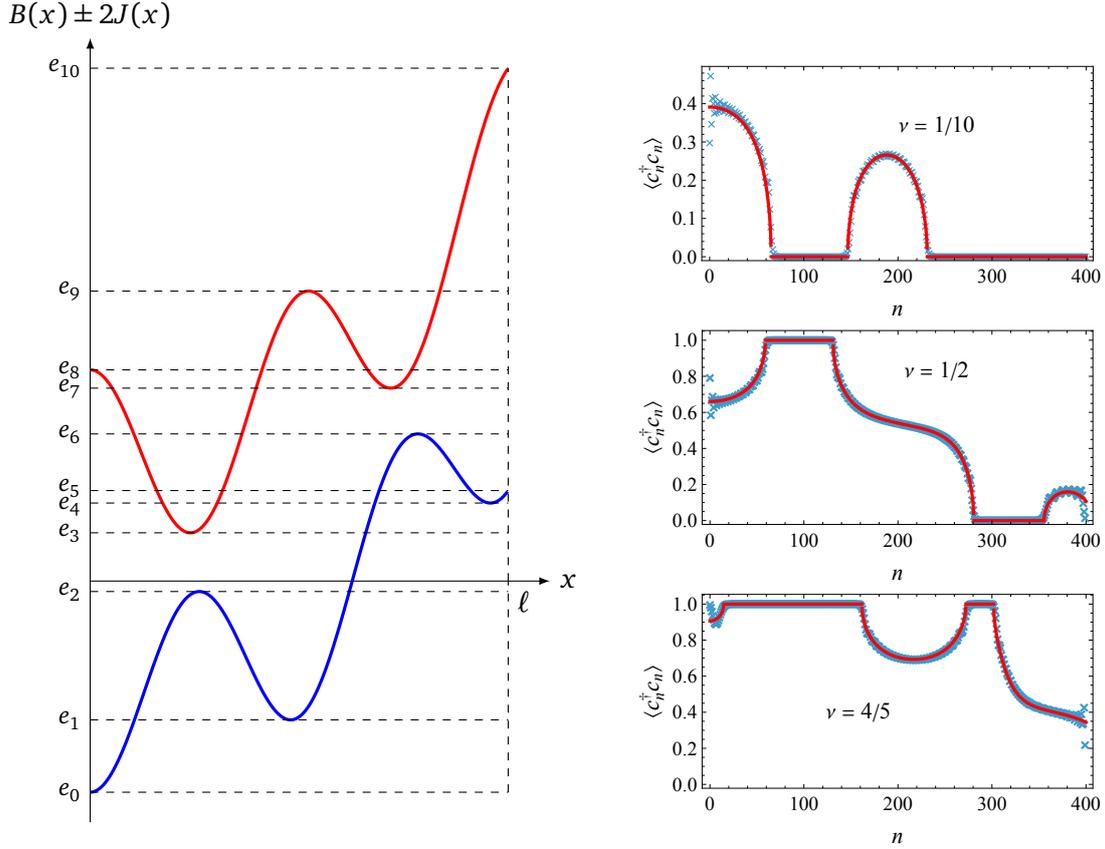%
\begin{table}[b!]
\centering
\caption{Critical values $e_i$ of the Fermi energy (with $i=1,\dots,9$) and their corresponding
  filling fraction $\nu_i$.}\smallskip
\begin{tabular}{|c|rrrrrrrrr|}
  \hline
  $e_i$& $-2.3009$ & $-0.1737$ & $0.7998$ & $1.2929$ & $1.5000$ & $2.4384$ & $3.1972$
  & $3.5000$ & $4.8055$\\ 
  \hline
  $\nu_i$ &  $0.0225$ & $0.2100$ & $0.3700$ & $0.4425$ & $0.4725$ & $0.6225$ & $0.7700$
  & $0.8225$ & $0.9350$\\
  \hline
\end{tabular}
\label{tab:nu_epsilon}
\end{table}%
More specifically, the behavior of $\rho(x,\vep_F)$ depends on the interval $[e_i,e_{i+1}]$
containing the Fermi energy $\vep_F$, where $e_j$ (with $j=0,1,\dots,10$) is one of the energies
plotted in fig.~\ref{fig:gen_cos_chain} (the numerical values of these energies and their
corresponding filling fractions are listed in table~\ref{tab:nu_epsilon}). For instance, if
$e_5<\vep_F<e_6$ (i.e., for filling fractions between $\nu_5=0.4725$ and $\nu_6=0.6225$) there are
four turning points $x_i(\vep_F)$ ($i=1,\dots,4$), with $x_1(\vep_F)>0$ and $x_4(\vep_F)<\ell$.
The fermionic density features a saturation interval $[x_1(\vep_F),x_2(\vep_F)]$ and a depletion
interval $[x_3(\vep_F),x_4(\vep_F)]$, while it varies continuously over the intervals
$[0,x_1(\vep_F)]$, $[x_2(\vep_F),x_3(\vep_F)]$, and $[x_4(\vep_F),\ell]$.

\begin{figure}[t!]
  \centering
  \includegraphics[height=.31\textwidth]{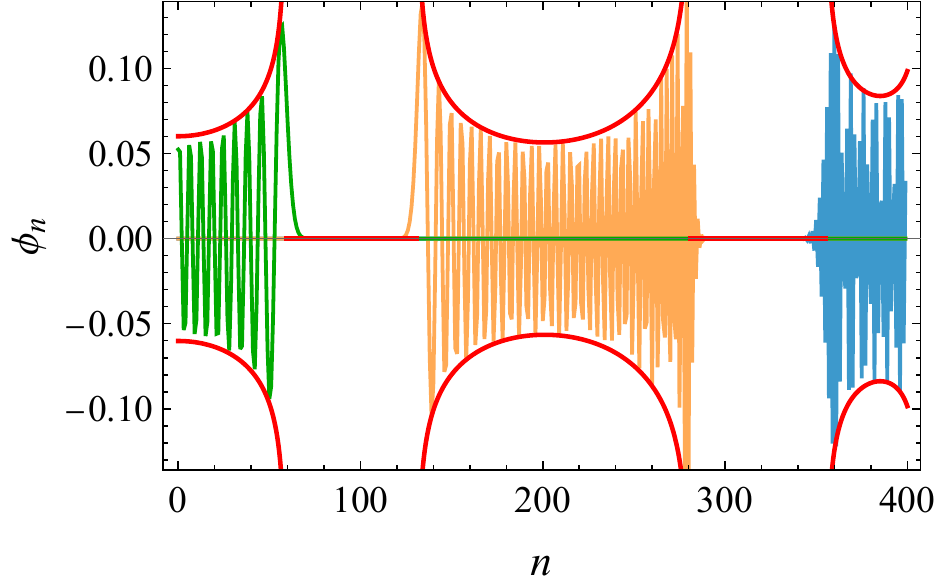}\hfill
  \includegraphics[height=.31\textwidth]{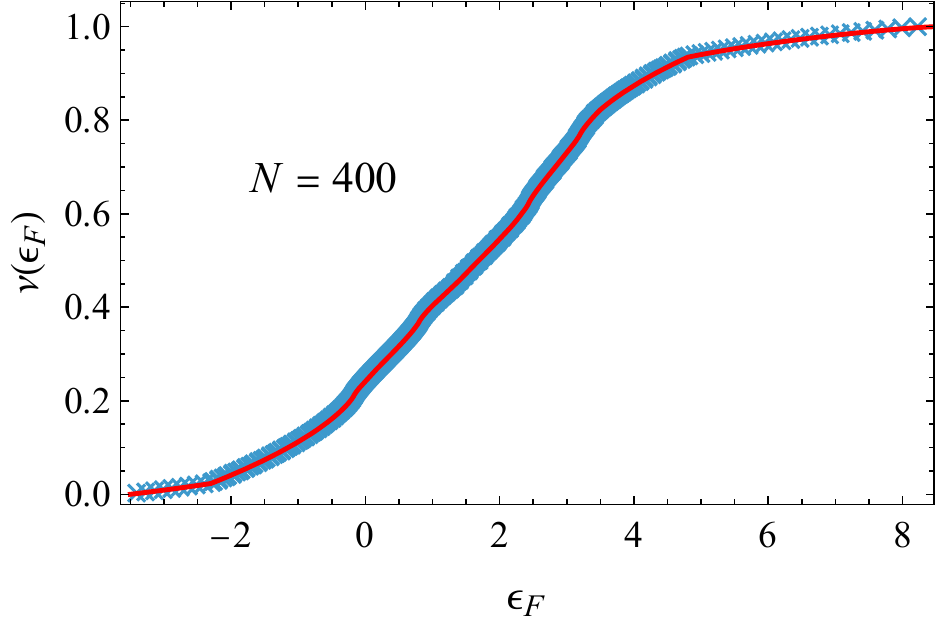}
  \caption{Left: linear combination~\eqref{eq:lin_comb} of the exact eigenfunctions of the
    chain~\eqref{eq:gen_cos_chain} with $N=400$ and energies $\vep_{N/2}$ (blue), $\vep_{N/2+1}$
    (orange) and $\vep_{N/2+2}$ (green), compared to its WKB envelope~\eqref{eq:envelope} (red
    curve). Right: filling fraction $\nu(\vep_F)$ as a function of the Fermi energy $\vep_F$ (blue
    markers) compared to its WKB approximation~\eqref{eq:filling} (red line).}
  \label{fig:phinu-gen-cos}
\end{figure}

As in the previous examples, the WKB approximations derived in the previous section are in
excellent agreement with the numerical results. For instance, in fig.~\ref{fig:phinu-gen-cos}
(left) we compare the WKB envelopes~\eqref{eq:envelope} with $\vep=\vep_{N/2}$ (for a chain with
$N=400$ spins) with the linear combination of three exact eigenfunctions with consecutive energies
$\vep_{N/2+i}$ ($i=0,1,2$), weighted as explained in remark~\ref{rem:eff_wf}. More precisely, from
fig.~\ref{fig:gen_cos_chain} and table~\ref{tab:nu_epsilon} it follows that for the energy
$\vep_{N/2}=1.69251\in(e_5,e_6)$ there are four turning points $x_1,\dots,x_4$ and three
potential wells $I_1=[0,x_1]$, $I_2=[x_2,x_3]$ and $I_3=[x_4,\ell]$. As illustrated in
fig.~\ref{fig:phinu-gen-cos} (left), the three exact eigenfunctions $\phi_n(\vep_{N/2})$,
$\phi_n(\vep_{N/2+1})$, and $\phi_n(\vep_{N/2+2})$ are respectively localized in the wells $I_3$,
$I_2$, and $I_1$. According to remark~\ref{rem:eff_wf} (see, in particular,
eq.~\eqref{eq:phi-phi_i}), eq.~\eqref{eq:envelope} with $\vep=\vep_{N/2}$ should provide a good
approximation to the envelope of the linear combination\footnote{Recall that we are taking $a=1$,
  and therefore $\ell=N$.}
\begin{equation}\label{eq:lin_comb}
  \sum_{i=1}^3\frac{A(\vep_{N/2})}{A_i(\vep_{N/2})}\,\phi_{n}(\vep_{N/2+i})
\end{equation}
(cf.~eq.~\eqref{eq:phi-phi_i}). This is in fact borne out by the figure.

It is also straightforward to obtain an estimate of the fermionic density for a given filling
fraction $\nu$ (or Fermi energy $\vep_F$) using eq.~\eqref{eq:occupation_density}. This formula is
surprisingly accurate over the whole range of filling fractions, as can be seen from
fig.~\ref{fig:gen_cos_chain} (right). Finally, by numerically integrating eq.~\eqref{eq:filling}
it is possible to derive an excellent asymptotic approximation to the filling fraction
$\nu(\vep_F)$ for arbitrary Fermi energy (see, e.g., fig.~\ref{fig:phinu-gen-cos} (right) for
$N=400$ spins).

\section{Conclusions and outlook}\label{sec:conc}

In this work, we developed a new analytical approach to study the fermion density profile of
inhomogeneous XX spin chains in the thermodynamic limit. By applying a WKB-like approximation
directly to the three-term recurrence relations satisfied by the single-particle eigenfunctions,
we constructed continuous approximations to these eigenfunctions without resorting to the
associated field theory. This led to a simple and general closed-form expression for the local
fermion density as a function of the Fermi energy, valid for arbitrary fillings, site-dependent
hopping amplitudes, and external magnetic fields.

Our density formula accurately captures both the depletion and saturation phenomena that were
previously observed numerically, but were not well understood analytically away from criticality
beyond the low-filling, zero-field regime. By testing the formula on a wide range of inhomogeneous
chains, including the Krawtchouk, rainbow, cosine, and (in the Supplementary Material) Rindler
chains, we demonstrated the excellent agreement of our asymptotic formula with the numerical
results even for moderately large systems.

In addition to its practical utility, our method may shed light on the mechanisms behind
entanglement suppression in inhomogeneous XX spin chains. For instance, in the simplest case in
which the eigenfunctions are localized in a single interval, the correlation matrix $C_{nm}$ can
be approximated in the thermodynamic limit as follows:
\begin{subequations}\label{eq:corr_mat}
  \begin{equation}\label{eq:Cnmapp}
    C_{nm}\equiv\mel{M}{c_n^\dagger c_m}{M}\to C(x,y)=a\int_{\vep_0}^{\vep_F}\dd\vep\, D(\vep)
    \phi(x,\vep)\phi(y,\vep) =\frac1\pi\,\int_{\vep_0}^{\vep_F}\dd\vep\, f(x,\vep)f(y,\vep),
  \end{equation}
  where $x=na$, $y=ma$, $\vep_F=\vep_{M-1}$, and
  \begin{equation}\label{eq:fs}
    f(s,\vep)=\frac{\sin\vp^*(s,\vep)}{\sqrt{J(s)\br(1-\xi(s,\vep)^2)^{1/2}}}\,
    \Th\!\br(1-\xi(s,\vep)^2).
  \end{equation}
\end{subequations}
Since the eigenstates of free-fermion systems are Gaussian~\cite{LR09}, the two-point correlation
matrix determines all higher-order correlations and entanglement properties. In particular, as
explained in section~\ref{sec:prelim}, the entanglement entropy of a subset
$A\subset\{0,\dots,N-1\}$ can be efficiently evaluated through eq.~\eqref{eq:entropy} in terms of
the eigenvalues $\nu_j$ of the truncated correlation matrix $\mathbf C_A=(C_{nm})_{n,m\in A}$. In
this way eqs.~\eqref{eq:corr_mat} could be the starting point for deriving asymptotic
approximations to the bipartite entanglement entropy for arbitrary fillings and external magnetic
fields, where the standard field-theoretic techniques based on conformal field
theory~\cite{HLW94,CC04JSTAT} are not directly applicable due to the lack of conformal invariance.
This could be of particular interest in the study of the asymptotic growth of the entanglement
entropy with the block size away from criticality, which in simple cases has been observed to
follow the usual area law for one-dimensional critical
systems~\cite{HLW94,CC04JSTAT,FC11,RDRCS17,TRS18} rescaled to the non-depletion
region~\cite{FG21}.

Finally, the discrete WKB approach introduced in this paper could also be applied in more general
scenarios, such as the construction of training densities incorporating non-local effects in the
evaluation of the exchange-correlation energy for certain fermion chains~\cite{LFARCB16}.

\section*{Acknowledgements}

The authors are grateful to the anonymous referees for their valuable comments on the first
version of the manuscript.


\paragraph{Funding information}
This work was partially supported by grants PID2024-155527NB-I00 from Spain's Mi\-nis\-te\-rio de
Ciencia, Innovación y Universidades and~PR12/24-31565 from Universidad Complutense de Madrid.




\section*{Supplementary Material}
An additional worked example presenting the computation of the fermionic density of the Rindler
chain for several values of its parameter is available at
\href{https://doi.org/10.5281/zenodo.17652257}{Zenodo repository}.









\begin{thebibliography}{10}
\providecommand{\url}[1]{\texttt{#1}}
\providecommand{\urlprefix}{URL }
\expandafter\ifx\csname urlstyle\endcsname\relax
  \providecommand{\doi}[1]{doi:\discretionary{}{}{}#1}\else
  \providecommand{\doi}{doi:\discretionary{}{}{}\begingroup
  \urlstyle{rm}\Url}\fi
\providecommand{\eprint}[2][]{\url{#2}}

\bibitem{GL14}
T.~Gra{\ss} and M.~Lewenstein,
\newblock \emph{Trapped-ion quantum simulation of tunable-range {H}eisenberg
  chains},
\newblock EPJ Quantum Technology \textbf{1}, 8(20) (2014),
\newblock \doi{10.1140/epjqt8}.

\bibitem{BMF21}
R.~E. Barfknecht, T.~Mendes-Santos and L.~Fallani,
\newblock \emph{Engineering entanglement {H}amiltonians with strongly
  interacting cold atoms in optical traps},
\newblock Phys. Rev. Res. \textbf{3}, 013112 (2021),
\newblock \doi{10.1103/PhysRevResearch.3.013112}.

\bibitem{RTLC17}
J.~Rodr\'{\i}guez-Laguna, L.~Tarruell, M.~Lewenstein and A.~Celi,
\newblock \emph{Synthetic {U}nruh effect in cold atoms},
\newblock Phys. Rev. A \textbf{95}, 013627 (2017),
\newblock \doi{10.1103/PhysRevA.95.013627}.

\bibitem{SFSTT20}
F.~Schäfer, T.~Fukuhara, S.~Sugawa, Y.~Takasu and Y.~Takahashi,
\newblock \emph{Tools for quantum simulation with ultracold atoms in optical
  lattices},
\newblock Nat. Rev. Phys. \textbf{2}, 411 (2020),
\newblock \doi{10.1038/s42254-020-0195-3}.

\bibitem{MCDetal21}
C.~Monroe, W.~C. Campbell, L.-M. Duan, Z.-X. Gong, A.~V. Gorshkov, P.~W. Hess,
  R.~Islam, K.~Kim, N.~M. Linke, G.~Pagano, P.~Richerme, C.~Senko
  \emph{et~al.},
\newblock \emph{Programmable quantum simulations of spin systems with trapped
  ions},
\newblock Rev. Mod. Phys. \textbf{93}, 025001 (2021),
\newblock \doi{10.1103/RevModPhys.93.025001}.

\bibitem{VRL10}
G.~Vitagliano, A.~Riera and J.~I. Latorre,
\newblock \emph{Volume-law scaling for the entanglement entropy in spin-$1/2$
  chains},
\newblock New J. Phys. \textbf{12}, 113049(16) (2010),
\newblock \doi{10.1088/1367-2630/12/11/113049}.

\bibitem{RDRCS17}
J.~Rodr{\'{i}}guez-Laguna, J.~Dubail, G.~Ram{\'{i}}rez, P.~Calabrese and
  G.~Sierra,
\newblock \emph{More on the rainbow chain: entanglement, space-time geometry
  and thermal states},
\newblock J. Phys. A: Math. Theor. \textbf{50}, 164001(18) (2017),
\newblock \doi{10.1088/1751-8121/aa6268}.

\bibitem{TRS18}
E.~Tonni, J.~Rodr{\'{i}}guez-Laguna and G.~Sierra,
\newblock \emph{Entanglement {H}amiltonian and entanglement contour in
  inhomogeneous 1{D} critical systems},
\newblock J. Stat. Mech.-Theory E. \textbf{2018}, 043105(39) (2018),
\newblock \doi{10.1088/1742-5468/aab67d}.

\bibitem{LSM61}
E.~Lieb, T.~Schultz and D.~Mattis,
\newblock \emph{Two soluble models of an antiferromagnetic chain},
\newblock Ann. Phys. \textbf{16}, 407 (1961),
\newblock \doi{10.1016/0003-4916(61)90115-4}.

\bibitem{FG20}
F.~Finkel and A.~Gonz\'alez-L\'opez,
\newblock \emph{Inhomogeneous {XX} spin chains and quasi-exactly solvable
  models},
\newblock J. Stat. Mech.-Theory E. \textbf{2020}, 093105(41) (2020),
\newblock \doi{10.1088/1742-5468/abb237}.

\bibitem{JW28}
P.~Jordan and E.~Wigner,
\newblock \emph{Über das {P}aulische \"{A}quivalenzverbot},
\newblock Z. Physik \textbf{47}, 631 (1928),
\newblock \doi{10.1007/BF01331938}.

\bibitem{VLRK03}
G.~Vidal, J.~I. Latorre, E.~Rico and A.~Kitaev,
\newblock \emph{Entanglement in quantum critical phenomena},
\newblock Phys. Rev. Lett. \textbf{90}, 227902(4) (2003),
\newblock \doi{10.1103/PhysRevLett.90.227902}.

\bibitem{Pe03}
I.~Peschel,
\newblock \emph{Calculation of reduced density matrices from correlation
  functions},
\newblock J. Phys. A: Math. Gen \textbf{36}, L205 (2003),
\newblock \doi{10.1088/0305-4470/36/14/101}.

\bibitem{LR09}
J.~I. Latorre and A.~Riera,
\newblock \emph{A short review on entanglement in quantum spin systems},
\newblock J. Phys. A: Math. Theor. \textbf{42}, 504002(33) (2009),
\newblock \doi{10.1088/1751-8113/42/50/504002}.

\bibitem{Re70}
A.~R{\'e}nyi,
\newblock \emph{Probability Theory},
\newblock North-Holland, Amsterdam (1970).

\bibitem{FG21}
F.~Finkel and A.~Gonz\'alez-L\'opez,
\newblock \emph{Entanglement entropy of inhomogeneous {XX} spin chains with
  algebraic interactions},
\newblock J. High. Energy Phys. \textbf{12}, 184(34) (2021),
\newblock \doi{10.1007/JHEP12(2021)184}.

\bibitem{MSSSR22}
B.~Mula, N.~Samos S{\'{a}}enz~de Buruaga, G.~Sierra, S.~N. Santalla and
  J.~Rodr{\'{i}}guez-Laguna,
\newblock \emph{Depletion in fermionic chains with inhomogeneous hoppings},
\newblock Phys. Rev. B \textbf{106}, 224204(10) (2022),
\newblock \doi{10.1103/PhysRevB.106.224204}.

\bibitem{RRS15}
G.~Ram{\'{i}}rez, J.~Rodr{\'{i}}guez-Laguna and G.~Sierra,
\newblock \emph{Entanglement over the rainbow},
\newblock J. Stat. Mech.-Theory E. \textbf{2015}, 06002(20) (2015),
\newblock \doi{10.1088/1742-5468/2015/06/P06002}.

\bibitem{CNV19}
N.~Cramp{\'{e}}, R.~I. Nepomechie and L.~Vinet,
\newblock \emph{Free-fermion entanglement and orthogonal polynomials},
\newblock J. Stat. Mech.-Theory E. \textbf{2019}, 093101(17) (2019),
\newblock \doi{10.1088/1742-5468/ab3787}.

\bibitem{BPV24}
G.~Blanchet, G.~Parez and L.~Vinet,
\newblock \emph{Fermionic logarithmic negativity in the {K}rawtchouk chain},
\newblock J. Stat. Mech.-Theory E. \textbf{2024}, 113101(24) (2024),
\newblock \doi{10.1088/1742-5468/ad84d8}.

\bibitem{BCLMV25}
P.-A. Bernard, N.~Cramp{\'e}, Q.~Labriet, L.~Morey and L.~Vinet,
\newblock \emph{Exactly solvable inhomogeneous {XY} spin chain},
\newblock J. Stat. Mech.-Theory E. \textbf{2025}, 103101(12) (2025),
\newblock \doi{10.1088/1742-5468/ae0429}.

\bibitem{DM67}
R.~B. Dingle and G.~J. Morgan,
\newblock \emph{{WKB} methods for difference equations {I}},
\newblock Appl. Sci. Res. \textbf{18}, 221 (1967),
\newblock \doi{10.1007/BF00382348}.

\bibitem{OLBC10}
F.~W.~J. Olver, D.~W. Lozier, R.~F. Boisvert and C.~W. Clark, eds.,
\newblock \emph{{NIST} Handbook of Mathematical Functions},
\newblock Cambridge University Press (2010).

\bibitem{Ge31}
S.~Gerschgorin,
\newblock \emph{{\"U}ber die {A}bgrenzung der {E}igenwerte einer {M}atrix},
\newblock Izv. Akad. Nauk SSSR Ser. Mat. \textbf{6}, 749 (1931).

\bibitem{EP18}
V.~Eisler and I.~Peschel,
\newblock \emph{Properties of the entanglement {H}amiltonian for finite
  free-fermion chains},
\newblock J. Stat. Mech.-Theory E. \textbf{2018}, 104001(17) (2018),
\newblock \doi{10.1088/1742-5468/aace2b}.

\bibitem{Fr58}
J.~Friedel,
\newblock \emph{Metallic alloys},
\newblock Nuovo Cimento Suppl. (1955--1965) \textbf{7}, 287 (1958),
\newblock \doi{doi.org/10.1007/BF02751483}.

\bibitem{BBEPV25}
P.~A. Bernard, R.~Bonsignori, V.~Eisler, G.~Parez and L.~Vinet,
\newblock \emph{Entanglement {H}amiltonian and orthogonal polynomials},
\newblock Nucl. Phys. B \textbf{1020}, 117185(16) (2025),
\newblock \doi{10.1016/j.nuclphysb.2025.117185}.

\bibitem{KLS10}
R.~Koekoek, P.~Lesky and R.~Swarttouw,
\newblock \emph{Hypergeometric {O}rthogonal {P}olynomials and their
  $q$-{A}nalogues},
\newblock Springer-Verlag, Berlin,
\newblock \doi{10.1007/978-3-642-05014-5} (2010).

\bibitem{RRS14}
G.~Ram{\'{i}}rez, J.~Rodr{\'{i}}guez-Laguna and G.~Sierra,
\newblock \emph{From conformal to volume law for the entanglement entropy in
  exponentially deformed critical spin $1/2$ chains},
\newblock J. Stat. Mech.-Theory E. \textbf{2014}, P10004(15) (2014),
\newblock \doi{10.1088/1742-5468/2014/10/P10004}.

\bibitem{MLNR19}
I.~MacCormack, A.~Liu, M.~Nozaki and S.~Ryu,
\newblock \emph{Holographic duals of inhomogeneous systems: the rainbow chain
  and the sine-square deformation model},
\newblock J. Phys. A: Mat. Theor. \textbf{52}, 505401(24) (2019),
\newblock \doi{10.1088/1751-8121/ab3944}.

\bibitem{Sz25}
A.~Szabó,
\newblock \emph{Rainbow chains and numerical renormalisation group for accurate
  chiral conformal spectra},
\newblock SciPost Phys. \textbf{19}, 075(30) (2025),
\newblock \doi{10.21468/scipostphys.19.3.075}.

\bibitem{HLW94}
C.~Holzhey, F.~Larsen and F.~Wilczek,
\newblock \emph{Geometric and renormalized entropy in conformal field theory},
\newblock Nucl. Phys. B \textbf{424}, 443 (1994),
\newblock \doi{10.1016/0550-3213(94)90402-2}.

\bibitem{CC04JSTAT}
P.~Calabrese and J.~Cardy,
\newblock \emph{Entanglement entropy and quantum field theory},
\newblock J. Stat. Mech.-Theory E. \textbf{2004}, P06002(27) (2004),
\newblock \doi{10.1088/1742-5468/2004/06/P06002}.

\bibitem{FC11}
M.~Fagotti and P.~Calabrese,
\newblock \emph{Universal parity effects in the entanglement entropy of {$XX$}
  chains with open boundary conditions},
\newblock J. Stat. Mech.-Theory E. \textbf{2011}, P01017(26) (2011),
\newblock \doi{10.1088/1742-5468/2011/01/P01017}.

\bibitem{LFARCB16}
M.~Lubasch, J.~I. Fuks, H.~Appel, A.~Rubio, J.~I. Cirac and M.-C. Bañuls,
\newblock \emph{Systematic construction of density functionals based on matrix
  product state computations},
\newblock New J. Phys. \textbf{18}, 083039(11) (2016),
\newblock \doi{10.1088/1367-2630/18/8/083039}.

\end{thebibliography}


\end{document}